\documentclass[aps, pre,amsmath,amssymb,twocolumn,floatfix]{revtex4}
\usepackage[usenames,dvipsnames]{color}
\usepackage{color}
\usepackage[pdftex]{graphicx}
\usepackage{epsfig}
\usepackage{verbatim}
\usepackage{soul}
\input epsf

\newcommand {\dr}{{\mathrm d}\mathbf{r}}

\newcommand {\rr}{\mathbf{r}}
\newcommand {\kk}{\mathbf{k}}

\newcommand{\matTT}[4]{\left(\hspace{-2mm}\begin{array}{cc} #1 & #2 \\ #3 & #4
			\end{array}\hspace{-2mm}\right)}

\begin{document}

\title{Solidification in soft-core fluids: disordered solids from fast solidification fronts}

\author{A.J.~Archer}
\email{A.J.Archer@lboro.ac.uk}
\affiliation{Department of Mathematical Sciences, Loughborough University, Loughborough LE11 3TU, UK}
\author{M.C.~Walters}
\email{M.Walters@lboro.ac.uk}
\affiliation{Department of Mathematical Sciences, Loughborough University, Loughborough LE11 3TU, UK}
\author{U. Thiele}
\email{u.thiele@uni-muenster.de}
\affiliation{Institut f\"ur Theoretische Physik, Westf\"alische Wilhelms-Universit\"at M\"unster, Wilhelm Klemm Str.\ 9, D-48149 M\"unster, Germany}
\affiliation{Center of Nonlinear Science (CeNoS), Westf\"alische Wilhelms Universit\"at M\"unster, Corrensstr. 2, 48149 M\"unster, Germany}
\author{E.~Knobloch}
\email{knobloch@berkeley.edu}
\affiliation{Department of Physics, University of California at Berkeley, Berkeley, CA 94720, USA}

\begin{abstract} 
Using dynamical density functional theory we calculate the speed of solidification fronts advancing into a quenched two-dimensional model fluid of soft-core particles. We find that solidification fronts can advance via two different mechanisms, depending on the depth of the quench. For shallow quenches, the front propagation is via a nonlinear mechanism. For deep quenches, front propagation is governed by a linear mechanism and in this regime we are able to determine the front speed via a marginal stability analysis. We find that the density modulations generated behind the advancing front have a characteristic scale that differs from the wavelength of the density modulation in thermodynamic equilibrium, i.e., the spacing between the crystal planes in an equilibrium crystal. This leads to the subsequent development of disorder in the solids that are formed. For the one-component fluid, the particles are able to rearrange to form a well-ordered crystal, with few defects. However, solidification fronts in a binary mixture exhibiting crystalline phases with square and hexagonal ordering generate solids that are unable to rearrange after the passage of the solidification front and a significant amount of disorder remains in the system.

\end{abstract} 
\maketitle 
\epsfclipon  


\section{Introduction}

The question of why some materials form a disordered glass rather than a crystalline solid when they are cooled or compressed is one of the most pressing questions in both physics and materials science. A glass, like a crystalline solid, has a yield stress, i.e., it responds like an elastic solid when subjected to stress below the yield stress. However, on examining the microscopic structure of a glass (quantified via a suitable two-point correlation function or structure function, such as the static structure factor $S(k)$, that can be measured in a scattering experiment \cite{HM}), one finds no real difference between the structure of the glass and the same material at a slightly higher temperature, when it is a liquid. In order to discern the difference between a glass and a liquid from examining the microscopic structure, one approach is to determine the {\em dynamic} structure function. In a liquid, the particles are able to rearrange themselves over time, so that their subsequent positions become decorrelated from their earlier locations. On the other hand, in a glass, the particle positions remain strongly correlated to their locations at an earlier time. The standard picture of this phenomenon is that the particles become trapped within a `cage' of neighboring particles so that in the glass the probability of a particle escaping is negligibly small \cite{HM}. Thus, in a glass the particles can be thought of as frozen in a disordered arrangement, instead of forming a periodic or crystalline lattice.

Much insight into the formation and the statistical and thermodynamic properties of glasses has been gained in recent years from the study of colloidal suspensions, because of our ability to observe and track individual colloids in suspension with a confocal microscope \cite{WeeksetalScience2000}. In this paper we investigate the structure and phase behavior of a simple two-dimensional (2D) model colloidal fluid composed of ultra-soft particles that are able to interpenetrate.  We first study the solidification of the one-component system, which generally forms a regular crystalline solid. We then investigate binary mixtures which form disordered solids much more readily, and relate the disorder we find to the solidification process when the system is quenched from the liquid state. In particular, we examine how solidification fronts propagate into the unstable liquid, and how this dynamical process can lead to disorder in the model \cite{ARTK12}. Our study of this system is based on density functional theory (DFT) \cite{Evans79, Evans92, HM, lutsko10} and dynamical density functional theory (DDFT) \cite{MaTa99, MaTa00, ArEv04, ArRa04}.

DFT is an obvious theoretical tool for studying the microscopic structure and phase behavior of confined fluids, because it provides a method for calculating the one-body (number) density $\rho(\rr)$ of a system confined in an external potential $\Phi(\rr)$. The density profile $\rho(\rr)$ gives the probability of finding a particle at position $\rr$ in the system and is obtained by minimizing the grand potential functional $\Omega[\rho]$ with respect to variations in $\rho(\rr)$ \cite{Evans92,HM}. Typically, this is done numerically, and one must discretise the density distribution $\rho(\rr) \to \rho_p$, recording it on a set of grid points (the index $p$ enumerates the grid points). One then numerically solves the discretised equations for the set $\{\rho_p\}$. An alternative approach is to assume the density profile $\rho(\rr)$ takes a specific functional form, parametrized by a set of parameters $\{\alpha_p\}$, and then seek the values of the parameters $\{\alpha_p\}$ minimizing the grand potential functional. This alternative technique is often used in studies of crystallization, where the density profile is (for example) assumed to be a set of Gaussian functions, centered on a set of lattice sites \cite{HM}.

Over the years, DFT has been used by several groups studying the properties of glassy systems. Wolynes and coworkers \cite{StWo84, SSW85, XiWo01} developed a successful model of hard-sphere glass formation based on the idea that the glass may be viewed as a system that is `frozen' onto a (random close-packed) nonperiodic lattice. This approach is based on the DFT theory for crystallization \cite{HM} and was followed up by a number of other investigations \cite{BaCo86, Lowen90, KaDa01, KaDa02, KiMu03, YYO07, CKDKS05, YOY08} extending and applying the method. All of these studies show that the free energy landscape may exhibit minima corresponding to the particles becoming localized (trapped) on a nonperiodic lattice. One limitation of these approaches is that the system is constrained by the choice of the nonperiodic lattice (or, in the case of the approach in Ref.\ \cite{YOY08}, by the fixed boundary particles). However, in the present work, rather than imposing a particular (nonperiodic) lattice structure on the system, we use DDFT to describe the solidification after the uniform liquid is deeply quenched to obtain the structure of the crystal or disordered solid that is formed as an {\em output}.

Here, we consider the case when the uniform fluid is quenched to a state point where the crystal is the equilibrium phase and we examine how the solid phase advances into the liquid phase, with dynamics described by DDFT. Our work here builds on earlier studies \cite{TBVL09, GaEl11, ARTK12} employing the phase field crystal (PFC) model \cite{ELWRGTG12} to explore a similar situation. The PFC free energy functional consists of a local gradient expansion approximation \cite{ARTK12, ELWRGTG12} and is arguably the simplest DFT that is able to describe both the liquid and crystal phases and the interface between them. In Refs.\ \cite{GaEl11, ARTK12} it was shown that the solidification front speed can be calculated by performing a marginal stability analysis, based on a dispersion relation obtained by linearising the DDFT (see Sec.\ \ref{sec:VA} below). The most striking result of the work in \cite{GaEl11, ARTK12} is the observation that the wavelength of the density modulations created behind such an advancing solidification front is not, in general, the same as that of the equilibrium crystal. Thus, for the system to reach the equilibrium crystal structure after such a solidification front passes through the system, significant rearrangements must occur and defects and disorder often remain. This conclusion, based on a marginal stability calculation in one dimension (1D), was confirmed in 2D PFC numerical simulations \cite{ARTK12}. In the present work we consider the same type of situation using a more sophisticated nonlocal DFT for fluids of soft penetrable particles. For this model fluid, we find that when the fluid is deeply quenched, the marginal stability analysis correctly predicts the solidification front speed, giving the same front speed as we obtain from direct numerical simulations. However, for shallow quenches we find that the front propagates via a nonlinear mechanism rather than the linear mechanism that underpins the marginal stability analysis, and that the 1D marginal stability analysis fails to predict the correct front speed. The overall picture that we observe is similar to that predicted for 2D systems on the basis of amplitude equations by Hari and Nepomnyashchy \cite{HN00}, as discussed further in the Appendix.

We also present results for a binary mixture of soft particles that exhibits several different competing crystal structures, including several hexagonal phases and a square phase. We find that when a solidification front advances through such a mixture a highly disordered state results, consisting of a patchwork of differently ordered regions, some that are square and others that are hexagonally coordinated. Thus, the solidification process generates disordered structures in a completely natural way.

This paper is structured as follows. In Sec.\ \ref{sec:model} we describe the model soft core fluids considered in this paper and briefly describe the Helmholtz free energy functional that we use as the basis of our DFT and DDFT calculations for the density profile(s) of the liquid and solid phases. In Sec.\ \ref{sec:fluid_structure} we examine the structure of the uniform fluid. We obtain and compare results for the radial distribution function $g(r)$, comparing results from a simple DFT that generates the Random Phase Approximation (RPA) closure to the Ornstein-Zernike equation with results from the Hyper-Netted Chain (HNC) closure approximation which is very accurate for soft systems, and find very good agreement between the two, thus validating the simple DFT that we use. In Sec.\ \ref{sec:phase_behavior} we present results for the equilibrium phase behavior of the one-component fluid, calculating the phase diagram. Then in Sec.\ \ref{sec:dynamics} we briefly describe the DDFT for the non-equilibrium fluid and calculate the dispersion relation for fluid mixtures. In Sec.\ \ref{sec:1comp_fronts} we briefly discuss the marginal stability analysis for determining front speeds and compare the results with those from 2D DDFT computations, and show that the solidifications fronts do not generate density modulations with the same wavelength as the equilibrium crystal. This leads to disorder, and we present results showing how the one-component system is able to rearrange over time to produce a well-ordered crystal, with only a few defects. In Sec.\ \ref{sec:bin_system} we present our results for a binary mixture of soft particles in which a solidification front can generate a solid with persistent disorder. Section \ref{sec:conc} contains concluding remarks. The Appendix describes an amplitude equation approach that helps explain the relation between the linear and nonlinear solidification fronts that we observe.

\section{Model fluid}
\label{sec:model}

In this paper we study 2D soft penetrable particles and their mixtures. We model the particles as interacting via the pair potential
\begin{equation}
v_{ij}(r)=\epsilon_{ij} e^{-(r/R_{ij})^n},
\label{eq:pair_pot}
\end{equation}
where the index $i,j=1,2$ labels particles of the two different species. The parameter $\epsilon_{ij}$ defines the energy for complete overlap of a pair of particles of species $i$ and $j$ and $R_{ij}$ defines the range of the interaction. We also consider a one-component fluid, and in this case omit the indices -- i.e., we write the interaction between the particles as $v(r)=\epsilon e^{-(r/R)^n}$. The case $n=2$ corresponds to the Gaussian core model (GCM) \cite{Stillinger76, BLHM01, LLWL00, Likos01} and larger values of $n$ define the so-called generalised exponential model of index $n$ (GEM-$n$). In this paper we focus on the cases $n=4$ and $n=8$. Penetrable spheres correspond to the limit $n\to\infty$. Such soft potentials provide a simple model for the effective interactions between polymers, star-polymers, dendrimers and other such soft macromolecules in solution \cite{Likos01, DaHa94, likos:prl:98, LBHM00, BLHM01, JDLFL01, Dzubiella_2001, LBFKMH02, likos:harreis:02, GHL04, MFKN05, Likos06, LBLM12}. For such particles one may approximate the intrinsic Helmholtz free energy of the system as \cite{Likos01}
 \begin{eqnarray}\notag
 \mathcal{F}[\{\rho_i(\rr)\}]=k_BT \sum_{i=1}^2\int \dr \rho_i(\rr)\left(\log[\rho_i(\rr)\Lambda_i^2]-1\right)\\
+ \frac{1}{2}\sum_{i,j}\int\dr\int\dr'\rho_i(\rr)v_{ij}(|\rr-\rr'|)\rho_j(\rr'),
 \label{eq:DFT}
 \end{eqnarray}
where $T$ is temperature, $k_B$ is Boltzmann's constant and $\Lambda_i$ is the (irrelevant) thermal de Broglie wavelength for species $i$. Henceforth we set $\Lambda_i=R_{11}=1$. The free energy is a functional of the one-body density profiles $\rho_1(\rr)$ and $\rho_2(\rr)$, where $\rr=(x,y)$. The first term in Eq.~\eqref{eq:DFT}, $\mathcal{F}_{id}$, is the ideal gas (entropic) contribution to the free energy while the second term $\mathcal{F}_{ex}$ is the contribution from the interactions between particles. The equilibrium density distribution is that which minimizes the grand potential functional
\begin{equation}
\Omega[\{\rho_i(\rr)\}]=\mathcal{F}[\{\rho_i(\rr)\}]+\sum_{i=1}^2\int\dr\rho_i(\rr)(\Phi_i(\rr)-\mu_i),
\label{eq:GPF}
\end{equation} 
where $\mu_i$ are the chemical potentials and $\Phi_i(\rr)$ is the external potential experienced by particles of species $i$. When evaluated using the equilibrium density profiles, the grand potential functional gives the thermodynamic grand potential of the system. For a system in the bulk fluid state (i.e., where $\Phi_i(\rr)=0$), the minimizing densities are independent of position, $\rho_i(\rr)=\rho_i^b$. However, at other state points, for example, when the system freezes to form a solid, $\Omega$ is minimised by nonuniform density distributions, exhibiting sharp peaks.

The free energy functional in Eq.~\eqref{eq:DFT} generates the random phase approximation (RPA) for the pair direct correlation functions,
\begin{equation}
c^{(2)}_{ij}(\rr,\rr')\equiv-\beta\frac{\delta^2 F_{ex}}{\delta \rho_i(\rr) \delta \rho_j(\rr')}=-\beta v_{ij}(|\rr-\rr'|),
\label{eq:RPA}
\end{equation} 
where $\beta\equiv 1/k_BT$. For three-dimensional (3D) systems of soft-particles such as those considered here, the simple approximation for the free energy in \eqref{eq:DFT} is known to provide a good approximation for the fluid structure and thermodynamics, as long as $\beta \epsilon$ is not too large and the density is sufficiently high, i.e., when the average density in the system $\rho R^2>1$ and the particles experience multiple overlaps with their neighbors -- the classic mean-field scenario \cite{Likos01}. Below, we confirm that this approximation is also good in 2D, by comparing results for the fluid structure with results from the more accurate HNC approximation. This simple DFT has been used extensively with great success to study the phase behavior and structure of soft particles and their mixtures \cite{ArEv01, ArEv02, AER02,  ALE02, ALE04, GAL06, MGKNL06, MGKNL07, MoLi07, LMGK07, MCLFK08, LMMGK08, OvLi09b, OvLi09, TMAL09, CML10, MaLi11, NKL12, CPPR12, Pini14}. However, the DFT in \eqref{eq:DFT} is unable to describe the solid phases of the GCM, i.e., GEM-2 -- in order to calculate the free energy and structure of the solid phases of the GCM, one must introduce additional correlation contributions to the free energy \cite{Archer05c}. In contrast, when $n>2$, the approximation in Eq.~\eqref{eq:DFT} is able to provide a good account of the free energy and structure of the solid phase in 2D whenever $\beta \epsilon\sim O(1)$ or smaller. Away from this regime, other approaches are needed \cite{MGKNL06, MCLFK08, ZCM10, ZC12, WS13, WS14}.

\section{Structure of the fluid}
\label{sec:fluid_structure}

The pair correlations in a fluid may be characterised by the radial distribution functions $g_{ij}(r)=1+h_{ij}(r)$, where $h_{ij}(r)$ are the fluid total correlation functions \cite{HM}. These are related to the fluid direct pair correlation functions $c_{ij}^{(2)}(r)$ via the Ornstein-Zernike equation, which for a binary fluid is
\begin{equation}
h_{ij}(r)=c_{ij}^{(2)}(r)+\sum^2_{k=1} \rho_{k} \int d \rr' c^{(2)}_{ik}(|\rr-\rr'|)h_{jk}(\rr').
\label{eq:OZ}
\end{equation} 	
This equation, together with the exact closure relation
\begin{equation}
c_{ij}^{(2)}(r)=-\beta v_{ij}(r)+h_{ij}(r)-\ln(h_{ij}(r)+1)+b_{ij}(r),
\label{eq:CR}
\end{equation}
may be solved for $h_{ij}(r)$ and hence $g_{ij}(r)$. However, the bridge functions $b_{ij}(r)$ in Eq.\ \eqref{eq:CR} are not known exactly and so approximations are required. For different interactions between particles, various approximations for $b_{ij}(r)$ have been developed \cite{HM}. For fluids of soft particles, the HNC approximation, which consists of setting $b_{ij}(r)=0$, has been shown to be very accurate \cite{Likos01}. Below, we compare the results for $g(r)$ for the one-component fluid, obtained from the HNC closure with those obtained from the simple approximate DFT in Eq.\ \eqref{eq:DFT}. These are obtained via the so-called ``test particle method'' which consists of fixing one of the particles in the fluid and then calculating the density profiles $\rho_{i}(r)$ in the presence of this fixed particle. One then uses the Percus result $g_{ij}(r)=\rho_i(r)/\rho^b$, where the fixed particle is of species $j$. The equilibrium fluid density profiles are those which minimise the grand free energy, i.e., they satisfy the Euler-Lagrange equations
\begin{equation}
\frac{\delta \Omega}{\delta \rho_i(\rr)}=0.
\label{eq:EL}
\end{equation}
From Eqs.\ \eqref{eq:DFT} and \eqref{eq:GPF} we obtain
\begin{equation}
k_BT\ln\rho_i+\sum_j \int d \rr' \rho_j(\rr')v_{ij}(|\rr-\rr'|)+\Phi_{i}(\rr)-\mu_i=0.
\label{eq:EL2}
\end{equation}
In the test particle situation, we set the external potentials equal to those corresponding to fixing one of the particles, i.e., $\Phi_{i}(\rr)=v_{ik}(r)$, for a fixed particle of species $k$. Using the conditions that as $r\rightarrow \infty$, $\Phi_{i}(\rr)\rightarrow 0$ and $\rho_i(\rr)\rightarrow \rho_i^b$, we can eliminate the chemical potentials $\mu_i$ from Eq.~\eqref{eq:EL2} and obtain
\begin{eqnarray}
k_BT\ln \left(\frac{\rho_i(r)}{\rho_i^b}\right)&+&\sum_j \int d \rr' (\rho_j(\rr')-\rho_j^b)v_{ij}(|\rr-\rr'|)\nonumber\\
&+&v_{ik}(r)=0.
\label{eq:NoPot}
\end{eqnarray}

\begin{figure}[h!]
	\includegraphics[width=1.\columnwidth]{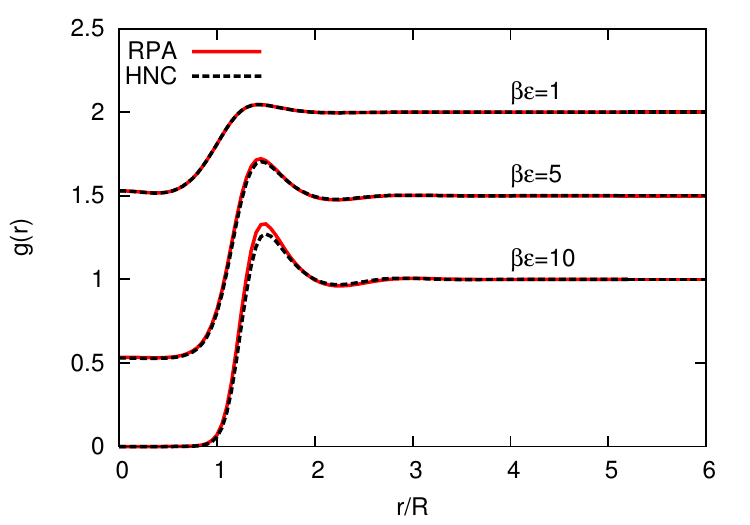}
	\caption{(Color online) The radial distribution function $g(r)$ for a GEM-4 fluid with bulk chemical potential $\mu=0$ obtained from the HNC closure to the OZ equation (dashed lines) and from the RPA DFT via the test particle method (solid lines), for several values of $\beta\epsilon$. For clarity, the results for $\beta\epsilon=1$ and 5 have been shifted vertically. The results correspond to the state points $(\beta\epsilon,\rho^bR^2)=(1,0.36)$, $(5,0.14)$ and $(10,0.088)$. As $\beta \epsilon$ increases, the RPA approximation becomes increasingly poor; nevertheless, even for (fairly low density) state points such as $\beta\epsilon=10$ the agreement is surprisingly good -- recall that the RPA approximation improves as the density is increased.}
	\label{RPAvHNC}
\end{figure}

We solve these equations using standard Picard iteration to obtain the density profiles $\rho_{ik}(r)$, where the index $k$ denotes the species held fixed. It is worth noting that if we replace the density profiles in Eq.~\eqref{eq:NoPot} by the total correlation functions, i.e., using $\rho_{ik}(r)=\rho_i^bg_{ik}(r)$, where $g_{ik}(r)=1+h_{ik}(r)$, we can rewrite Eq.~\eqref{eq:NoPot} in the form 
\begin{equation}
h_{ik}(r)=c_{ik,HNC}^{(2)}(r)+\sum_j \rho_j^b \int d \rr' h_{ij}(\rr')c_{ij,RPA}^{(2)}(\rr-\rr')
\label{eq:TP}
\end{equation}
(cf.~Eq.~\eqref{eq:OZ}), where $c_{ij,HNC}^{(2)}(r)$ denotes the HNC closure approximation for the pair direct correlation function (i.e., setting $b_{ij}(r)\equiv 0$ in Eq.~\eqref{eq:CR}) and $c_{ij,RPA}^{(2)}(r)=-\beta v_{ij}(r)$ denotes the RPA approximation. In Fig.~\ref{RPAvHNC} we compare results from the HNC closure of the OZ equation and the RPA test particle results for a one-component fluid with chemical potential $\mu=0$ and various values of $\beta \epsilon$. We see that the agreement between the two is very good, even at low temperatures such as $\beta\epsilon=10$, where one might expect the RPA to fail.

\section{Equilibrium fluid phase behavior}
\label{sec:phase_behavior}

\begin{figure}[t]
   \includegraphics[width=1.\columnwidth]{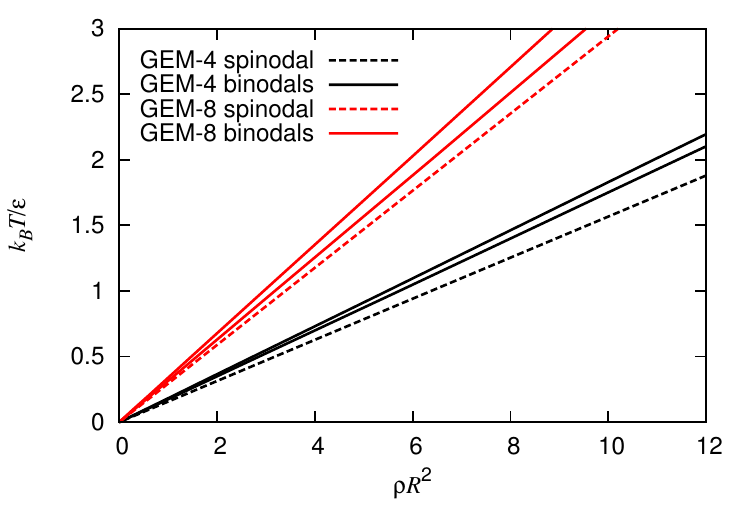}
   \caption{(Color online) Phase diagrams of the one-component 2D GEM-4 and GEM-8 model fluids. The solid lines are the binodals, i.e., loci of coexisting liquid and solid phases. The dashed lines are the spinodal-like instability lines along which the metastable liquid phase becomes linearly unstable.}
   \label{fig:phase_diag}
\end{figure}

\begin{figure}[t]
\includegraphics[width=0.9\columnwidth]{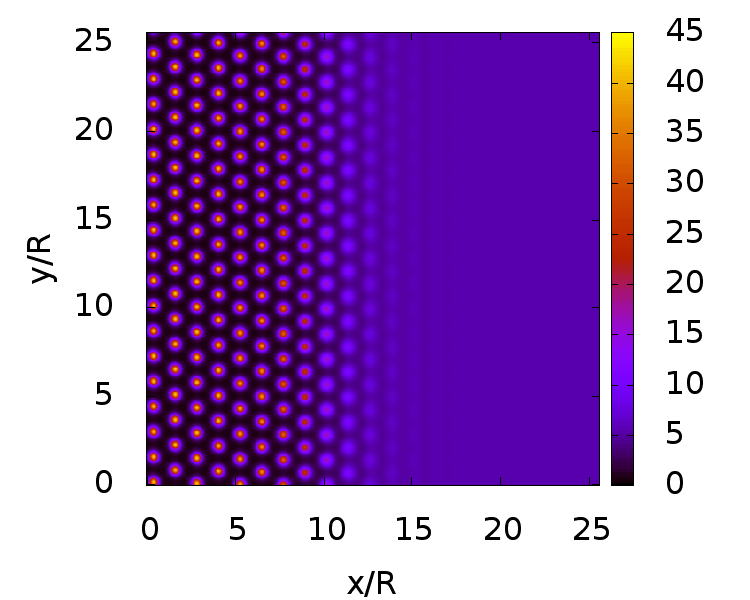} 
\caption{(Color online) Equilibrium density profile at the free interface between coexisting liquid and solid phases in the GEM-4 model when $\beta \epsilon=1$ and $\beta\mu=17.0$.}
\label{fig:free_interface}
\end{figure}

Having established that the simple RPA approximation for the free energy \eqref{eq:DFT} gives a good description of the structure of the bulk fluid, we now apply this to determine the phase diagram of the one-component GEM-4 and GEM-8 models -- in particular, to determine where the fluid freezes to form a crystal. We calculate the density profile of the uniform solid by solving the Euler-Lagrange equation \eqref{eq:EL} using a simple iterative algorithm on a 2D discretised grid with periodic boundary conditions. The uniform density system is linearly unstable at higher densities (this notion is discussed further below) and so for these state points it is easy to calculate the density of the crystal phase. An initial condition consisting of a line along which the density is higher than elsewhere, plus an additional small random number to break the symmetry of the profile, is sufficient. The density profile of the crystal obtained at higher densities is then continued down to lower densities where the liquid and crystal phases coexist.

Two phases coexist when the temperature, pressure and chemical potential of the two phases are equal. The densities of the coexisting liquid and crystal states in the 2D GEM-4 and GEM-8 models are displayed as a function of temperature in Fig.\ \ref{fig:phase_diag}. Qualitatively, the phase diagram is very similar to that found previously for the system in three dimensions (3D) \cite{MGKNL06, MGKNL07, LMGK07}. However, in the 2D case there is only one solid phase, unlike in 3D, where the system can form both fcc and bcc crystals, depending on the state point. The GEM-4 particles freeze at a higher density than the GEM-8 particles, because the GEM-4 potential is softer.

In Fig.\ \ref{fig:free_interface} we display a plot of the equilibrium density profile for the interface between the [1,1] crystal surface and the liquid. This density profile is for the GEM-4 model at temperature $\beta \epsilon=1$. At this temperature the chemical potential at coexistence is $\beta\mu=17$ and the densities of the coexisting liquid and solid phases are $\rho_{\rm l}R^2=5.48$ and  $\rho_{\rm s}R^2=5.73$, respectively. Similar stationary fronts at present even when the two competing phases are not exactly at thermodynamic coexistence -- a consequence of pinning of the front to the hexagonal structure to its left.

 \section{Theory for the non-equilibrium system}
 \label{sec:dynamics}
 
To extend the theory to non-equilibrium conditions, we assume the particles obey Brownian dynamics, modelled via overdamped stochastic equations of motion:
\begin{equation}
\dot{\rr}_l= -\Gamma_l\nabla_l U(\{\rr_l\},t) + \Gamma_l{\bf X}_l(t).
\label{eq:EOM}
\end{equation}
Here the index $l=1,..,N$ labels the particles, with $N\equiv N_1+N_2$ the total number of particles in the system and $N_i$ the number of particles of species $i$. The potential energy of the system is denoted by $U(\{\rr_l\},t)$, $\nabla_l\equiv\partial/\partial\rr_l$, ${\bf X}_l(t)$ is a white noise term and the friction constant $\Gamma_l^{-1}$ takes one of two values, $\Gamma_1^{-1}$ or $\Gamma_2^{-1}$, depending on the particle species. The quantities $\Gamma_i^{-1}$ characterise the drag of the solvent on particles of species $i$. The dynamics of a fluid of Brownian particles can be investigated using DDFT \cite{MaTa99,MaTa00,ArEv04,ArRa04}, which builds upon equilibrium DFT and takes as input the equilibrium fluid free energy functional. The two-component generalization of DDFT takes the form \cite{Archer05, Archer05b}
 \begin{equation}
 \frac{\partial\rho_i(\rr,t)}{\partial t} = 
  \Gamma_i \nabla\cdot\left[\rho_i(\rr,t)\nabla\frac{\delta\Omega[\{\rho_i(\rr,t)\}]}{\delta\rho_i(\rr,t)}\right],
 \label{eq:DDFT}
 \end{equation}
where $\rho_i(\rr,t)$ are now the time-dependent non-equilibrium fluid one-body density profiles. To derive the DDFT we use the approximation that the non-equilibrium fluid two-body correlations are the same as those in the equilibrium fluid with the same one-body density distributions \cite{MaTa99, MaTa00, ArEv04, ArRa04}.

\subsection{Fluid structure and linear stability}
\label{sec:VA}

We first consider the stability properties of a uniform fluid with densities $\rho_1^b$ and $\rho_2^b$, following the presentation in Refs.~\cite{ArEv04,ARTK12} (see also \cite{Evans79,Evans:TDGammaMolecP1979}). We set the external potentials $\Phi_i(\rr,t)=0$ and consider small density fluctuations $\tilde{\rho}_i(\rr,t)=\rho_i(\rr,t)-\rho_i^b$ about the bulk values. From Eq.~\eqref{eq:DDFT} we obtain
\begin{eqnarray}\notag
\frac{\beta}{\Gamma_i} \frac{\partial \tilde{\rho}_i(\rr,t)}{\partial t}
\, =\, \nabla^2 \tilde{\rho}_i(\rr,t)
\,-\, \rho_i^b \nabla^2 c_i^{(1)}(\rr,t) \\
-\, \nabla . [ \, \tilde{\rho}_i(\rr,t) \nabla c_i^{(1)}(\rr,t) \, ],
\label{eq:DDFT_2}
\end{eqnarray}
where
\begin{equation}
c_i^{(1)}(\rr)\equiv-\beta \frac{\delta (\mathcal{F}-\mathcal{F}_{id})}{\delta \rho_i(\rr)}
\end{equation}
are the one-body direct correlation functions \cite{Evans79,Evans92}. Taylor-expanding $c_i^{(1)}$ about the bulk values gives
\begin{equation}
c_i^{(1)}(\rr)=c_i^{(1)}(\infty)
+\sum_{j=1}^2\int \dr' \frac{\delta c_i^{(1)}(\rr)}{\delta
\rho_j(\rr')} \Bigg \vert_{\{\rho_i^b\}} \tilde{\rho}_j(\rr',t)+O (\tilde{\rho}^2),
\label{eq:c_1_expansion}
\end{equation}
where $c_i^{(1)}(\infty) \equiv c_i^{(1)}[\{\rho_i^b\}]=-\beta \mu_{i,ex}$ and $\mu_{i,ex}$ is the bulk excess chemical potential of species $i$. Since $\frac{\delta c_i^{(1)}(\rr)}{\delta \rho_j(\rr')}= c_{ij}^{(2)}(\rr,\rr')$, Eq.\ \eqref{eq:DDFT_2} yields, to linear order in $\tilde{\rho}_i$,
\begin{eqnarray}\notag
\frac{\beta}{\Gamma_i} \frac{\partial \tilde{\rho}_i(\rr,t)}{\partial t}
\, =\, \nabla^2 \tilde{\rho}_i(\rr,t)\hspace{2.5cm} \\
-\sum_j \rho_i^b \nabla^2 [ \, \int \, \dr' c_{ij}^{(2)}(|\rr-\rr'|)
\tilde{\rho}_j(\rr',t) \, ].
\label{eq:DDFT_linear}
\end{eqnarray}
A spatial Fourier transform of this equation yields an equation for the time evolution of the Fourier transform $\hat{\rho}_j(\kk,t)=\int \dr \exp({\rm i} \kk.\rr) \tilde{\rho}_j(\rr,t)$, where ${\rm i}=\sqrt{-1}$. We obtain
\begin{equation}
\frac{\beta}{\Gamma_i} \frac{\partial \hat{\rho}_i(k,t)}{\partial t}
=-k^2 \hat{\rho}_i(k,t)
+\rho_i^b\sum_j k^2 \, \hat{c}_{ij}(k) \hat{\rho}_j(k,t),
\label{eq:DDFT_linear_FT}
\end{equation}
where $\hat{c}_{ij}(k)\equiv\int \dr \exp({\rm i} \kk.\rr) c^{(2)}_{ij}(r)$ is the Fourier transform of the pair direct correlation function. If we assume that the time dependence of the Fourier modes follows $\hat{\rho}_i(k,t)\propto\exp[\omega(k)t]$ we obtain \cite{PBMT05, RAT11,RATK12,LK13}
\begin{equation}		
	\mathbf{1}\omega(k) \hat{\rho} = \mathbf{M} \cdot \mathbf{E} \hat{\rho} ,
	\label{eqMatrixForm}
\end{equation}
where $\hat{\rho}\equiv (\hat{\rho}_1,\hat{\rho_2})$ and the matrices $\mathbf{M}$ and $\mathbf{E}$ are given by
\begin{eqnarray}
	\mathbf{M} &=& \matTT{-k_BT\Gamma_1\rho_1^b k^2}{\hspace{5mm}0}{0}
	{-k_BT\Gamma_2\rho_2^b k^2}, \\ 
	\mathbf{E} &=& \matTT{\,\left[\frac{1}{\rho_1^b}-\hat{c}_{11}(k)\right]}{-\hat{c}_{12}(k)}
	{-\hat{c}_{21}(k)}{\left[\frac{1}{\rho_2^b}-\hat{c}_{22}(k)\right]\,}.
\end{eqnarray}
It follows that
%
\begin{equation}
	\omega(k) = \frac{1}{2}\mbox{Tr}(\mathbf{M \cdot E}) \pm 
	\sqrt{\frac{1}{4}\mbox{Tr}(\mathbf{M \cdot E})^2 - |\mathbf{M \cdot E}|}.
	\label{eqBetaTwoComp}
\end{equation}
where $|\mathbf{M \cdot E}|$ denotes the determinant of the matrix $\mathbf{M \cdot E}$.
When $\omega(k)<0$ for all wave numbers $k$, the system is linearly stable. If, however, $\omega(k)>0$ for any wave number $k$ then the uniform liquid is linearly unstable. Since $\mathbf{M}$ is a (negative definite) diagonal matrix its inverse $\mathbf{M^{-1}}$ exists for all nonzero densities and temperatures, enabling us to write Eq.\ \eqref{eqMatrixForm} as a generalised eigenvalue problem:
%
\begin{equation}
	(\mathbf{E} - \mathbf{M^{-1}} \omega) \hat{\rho} = 0.
\end{equation}
As $\mathbf{E}$ is a symmetric matrix, all eigenvalues are real as one would expect for a relaxational system. It follows that the threshold for linear instability is determined by $|\mathbf{E}| = 0$, i.e., by the condition
\begin{equation}
D(k)\equiv [1-\rho_1^b\hat{c}_{11}(k)][1-\rho_2^b\hat{c}_{22}(k)]-\rho_1^b\rho_2^b\hat{c}_{12}^2(k)=0.
\label{eq:lin_instab_crit}
\end{equation} 
The partial structure factors $S_{ij}(k)$ for an equilibrium fluid mixture are given by \cite{HM, ArEv01, ALE02, ALE04, BhTh70, Salmon06}
\begin{eqnarray}
\nonumber
      S_{11}(k) & = & 1 + \rho_1^b{\hat h}_{11}(k),\\
\nonumber
      S_{22}(k) & = & 1 + \rho_2^b{\hat h}_{22}(k),\\
      S_{12}(k) & = & \sqrt{\rho_1^b\rho_2^b}{\hat h}_{12}(k),
\end{eqnarray}
where ${\hat h}_{ij}(k)$ are the Fourier transforms of $h_{ij}(r)$, i.e., of the fluid pair correlation functions. These are related to the pair direct correlation functions $c_{ij}^{(2)}(r)$ through the Ornstein-Zernike (OZ) equations \cite{HM,ArEv01}. In Fourier space the OZ equations are
\begin{equation}
{\hat{h}}_{ij} (k) \, =\, \frac{N_{ij}(k)}{D(k)},
\end{equation}
with the three numerators given by
\begin{eqnarray}
N_{11} (k) &=& {\hat{c}}_{11}(k)+ \rho_2^b[{\hat{c}}_{12}^2(k)-
{\hat{c}}_{11}(k){\hat{c}}_{22}(k)]\notag \\
N_{22} (k) &=& {\hat{c}}_{22}(k)+ \rho_1^b[{\hat{c}}_{12}^2(k)-
{\hat{c}}_{11}(k){\hat{c}}_{22}(k)]\notag \\
N_{12} (k) &=& {\hat{c}}_{12}(k).
\end{eqnarray}
Since for an equilibrium fluid $S_{11}(k)>0$, $S_{22}(k)>0$ and $S_{11}S_{22}-S_{12}^2>0$ for all values of $k$, it follows that $D(k)>0$ and hence that $\omega(k)<0$ for all wave numbers $k$. Thus all Fourier modes decay over time. Within the present RPA theory for GEM-$n$ particles $\hat{c}_{ij}(k)=-\beta \hat{v}_{ij}(k)$, where $\hat{v}_{ij}(k)$ are the Fourier transforms of the pair potentials in Eq.~\eqref{eq:pair_pot}, and for sufficiently high densities $D(k)$ dips below zero. Thus $\omega(k)>0$ for a band of wave numbers around $k\approx k_c$, indicating that the fluid has become linearly unstable.

For a one-component fluid, i.e., in the limit of $\rho_2^b\to 0$, we find that the fluid is stable when $[1-\rho^b\hat{c}(k)]>0$ but becomes linearly unstable when $[1-\rho^b\hat{c}(k)]<0$ \cite{ArEv04, ARTK12}. The loci $D(k=k_c)=0$ for both the GEM-4 and GEM-8 models are displayed as dashed lines in Fig.~\ref{fig:phase_diag}. In both cases the line along which the liquid phase becomes linearly unstable is located well inside the region where the crystal is the equilibrium phase.

\section{Solidification fronts in the one-component GEM-4 model}
\label{sec:1comp_fronts}

\begin{figure}[t]
   \includegraphics[width=1.\columnwidth]{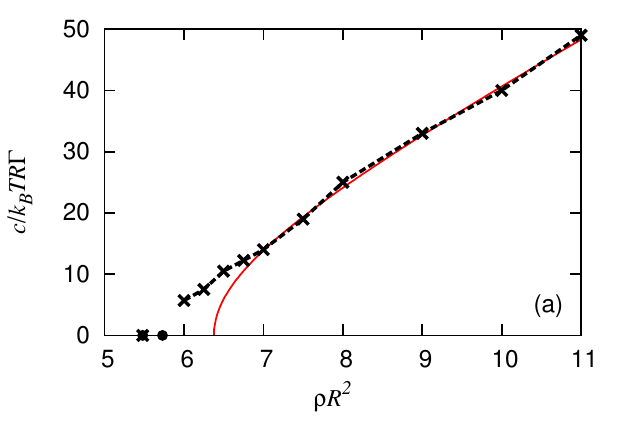}
   \includegraphics[width=1.\columnwidth]{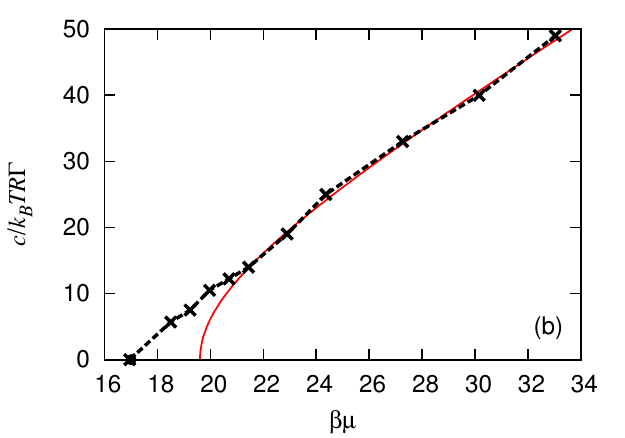}
\caption{\label{fig:speed} (Color online) The front speed (a) as a function of the density of the metastable liquid into which the front propagates and (b) as a function of the chemical potential, for a GEM-4 fluid with temperature $k_BT/\epsilon=1$. The red solid line is the result from the marginal stability analysis and the black dashed line is the result from numerical computations from profiles such as that displayed in Fig.\ \ref{fig:front_profile}. The black circles denote (a) the densities $\rho_{\rm l}$, $\rho_{\rm s}$ at liquid-solid coexistence, and (b) the coexistence value $\beta \mu\approx17.0$.}
\end{figure}

\begin{figure*}
\includegraphics[width=0.99\textwidth]{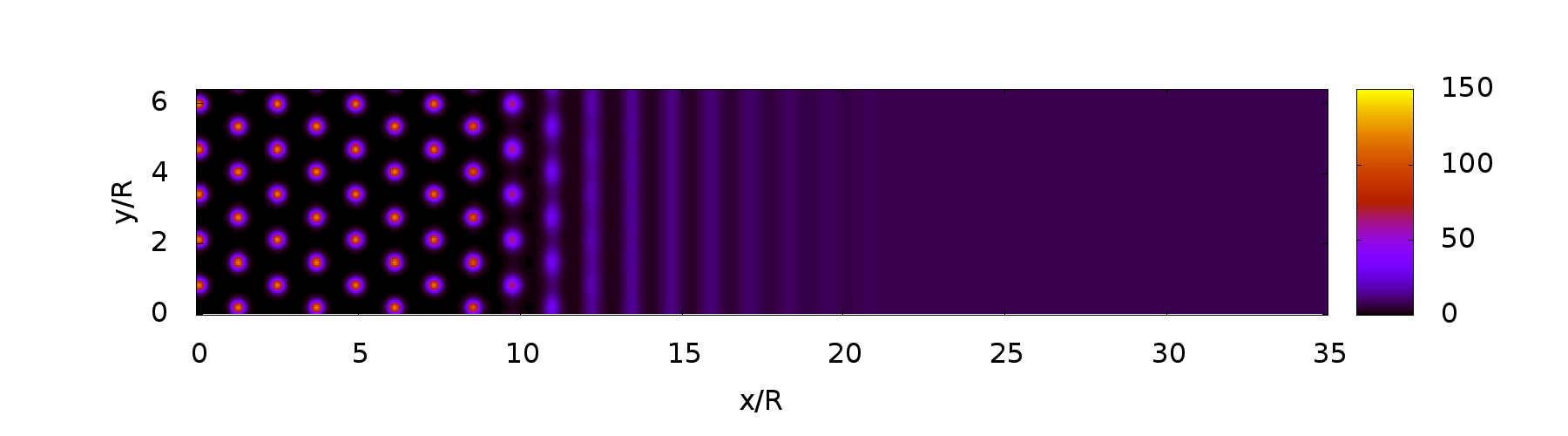}
 
 \includegraphics[width=0.9\textwidth]{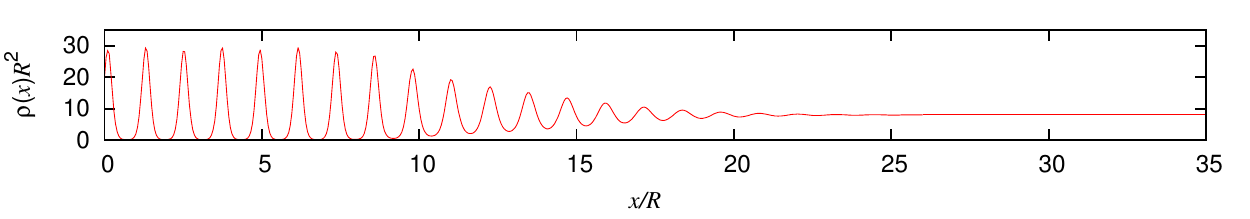}

\includegraphics[width=0.9\textwidth]{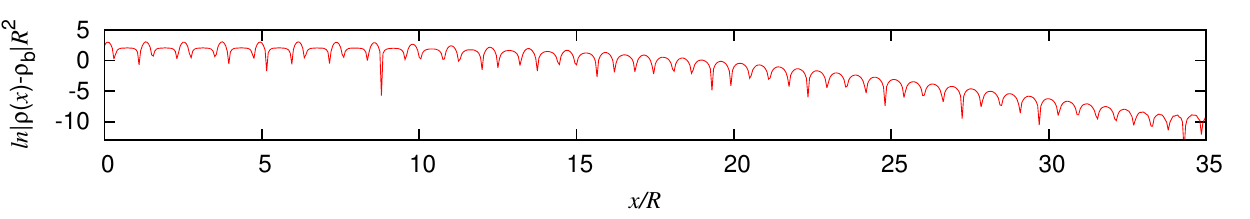}

\caption{\label{fig:front_profile} (Color online) Density profile across a solidification front advancing from left to right into an unstable GEM-4 liquid with bulk density $\rho R^2=8$ and temperature $k_BT/\epsilon=1$, calculated from DDFT. The top panel shows the full 2D density profile $\rho(x,y)$ while the panel below shows the 1D density profile $\rho(x)$ obtained by averaging over the $y$-direction, perpendicular to the front. The bottom panel shows $\ln(|\rho(x)-\rho^b|R^2)$ in order to reveal the small amplitude oscillations at the leading edge of the advancing front.}
\end{figure*}

When the system is linearly unstable, any localised density modulation will grow and advance into the unstable uniform liquid phase. In Refs.\ \cite{GaEl11, ARTK12}, a marginal stability analysis was used to calculate the speed of such a front for the PFC model. Such a calculation allows one to obtain the speed of a front that has advanced sufficiently far for all initial transients to have decayed, so that the front attains a stationary front velocity. In 1D the speed $c$ with which the front advances into the unstable liquid may be obtained by solving the following set of equations \cite{DL83, BBDKL85, GaEl11, ARTK12}:
\begin{eqnarray}\label{eq:speed1}
{\rm i}c+\frac{d \omega(k)}{dk}=0\\
{\rm Re}[{\rm i}ck+\omega(k)]=0,
\label{eq:speed2}
\end{eqnarray}
corresponding to a front solution moving with speed $c$ that is marginally stable to infinitesimal perturbations in its frame of reference. In such a front the density profiles are coupled (via the solution of the linear problem (\ref{eqMatrixForm})) and both take the form $\tilde{\rho}(\rr,t) = \rho_{\rm front}(x-ct)$, where $\rho_{\rm front}(x-ct) \sim \exp(-k_{\rm im}x)\sin(k_{\rm r} (x-ct)+{\rm Im}[\omega(k)]t)$. Here $k_{\rm r}$ and $k_{\rm im}$ are the real and imaginary parts of the complex wave number $k\equiv k_{\rm r}+{\rm i}k_{\rm im}$. The speed calculated from this approach for the one-component GEM-4 model is displayed as the solid red line in Fig.\ \ref{fig:speed}(a) as a function of the density of the unstable liquid and in (b) as a function of the chemical potential $\mu$, both for $\beta \epsilon=1$. We also display the front speed calculated numerically using DDFT in 2D. Figure \ref{fig:front_profile} shows typical 2D and 1D density profiles used for determining the front speed $c$. The figure shows that the invasion of the metastable liquid state in fact occurs via a {\it pair} of fronts, the first of which describes the invasion of the liquid state by an unstable pattern of stripes, while the second describes the invasion of the unstable stripe pattern by a stable hexagonal state. By ``stripes'' we mean a density profile with oscillations perpendicular to the front, but no density modulations parallel to the front. This double front structure complicates considerably the description of the invasion process in 2D (see Appendix). Figure \ref{fig:speed} shows measurements of the speed of propagation of the hexagons-to-stripes front, obtained by comparing profiles like that in Fig.\ \ref{fig:front_profile}(a) at two successive times and determining the speed of advance of the hexagonal state when it first emerges from the unstable stripe state. The speed of the stripe pattern is harder to measure since the pattern is itself unstable and so never reaches a substantial amplitude. For this reason we measure the speed of the stripe-to-liquid front from plots of the logarithm of the density fluctuations (Fig.~\ref{fig:front_profile}(c)) which emphasizes the spatial growth of the smallest fluctuations at the leading edge of the front.

For $\beta\epsilon=1$ the uniform liquid is linearly stable for $\beta\mu\lesssim 19.6$ and unstable for $\beta\mu\gtrsim19.6$. The marginal stability prediction, obtained by solving Eqs.~(\ref{eq:speed1}) and (\ref{eq:speed2}), predicts that the 1D speed increases with $\beta\mu$ (or with increasing density $\rho$) in a square-root manner, as indicated by the solid red line in Fig.~\ref{fig:speed}. Since the theory is 1D this prediction applies to the invasion of the liquid state by the stripe pattern. Despite this we find that the prediction correctly describes the speed of the hexagons-to-stripes front for $\beta\mu \gtrsim21.5$ (i.e.\ for $\rho R^2 \gtrsim7$), as measured in numerical simulations of the DDFT for the GEM-4 fluid, suggesting that the two fronts are locked together and that the front speed is selected by linear processes at the stripe-to-liquid transition, i.e., the resulting double front is a {\it pulled} front \cite{saarloos}. For smaller values of $\beta\mu$ the speed of the hexagonal state departs substantially from the marginal stability prediction and the stripe section is swallowed by the faster moving hexagons-to-liquid front. Indeed, for $\beta\mu\lesssim 19.6$ (i.e.\ for $\rho R^2\lesssim 6.38$) the stripe state is absent altogether, as can be verified by performing a parallel study in one spatial dimension. The bifurcation to stripes is therefore supercritical. The hexagons-to-liquid front present in the metastable regime below the onset of linear instability of the liquid state is stationary at the Maxwell point at $\beta\mu\approx 17.0$, corresponding to the location of thermodynamic coexistence between the liquid and hexagonal states. For $\beta\mu > 17.0$ the hexagonal state advances into the liquid phase (the opposite occurs for $\beta\mu < 17.0$) and the hexagons-to-liquid front is {\it pushed} \cite{saarloos}: in this regime the front propagates via a nonlinear process since the liquid phase is linearly stable. The situation is more subtle when plotted as a function of the liquid density $\rho R^2$: when the liquid density takes a value in the interval $5.48\lesssim \rho R^2\lesssim 5.73$, i.e., between the densities of the liquid and crystalline states at coexistence, one cannot define a unique front speed. In this regime any front between these two states will slow down and, in any finite domain, eventually come to a halt. This occurs because the density $\rho_0$ of the liquid state into which the front moves is less than the density $\rho_\mathrm{s}$ of the crystal at coexistence but larger than the density $\rho_\mathrm{l}$ of the liquid at coexistence. In this situation, the moving `front' has a substructure consisting of two transitions: one from $\rho_\mathrm{s}$ to a depletion zone of a density close to $\rho_\mathrm{l}$ and another one from the depletion zone to the initial $\rho_0$. As the depletion zone widens in time and limits the diffusion from the region of density $\rho_0$ to the crystalline zone of density $\rho_\mathrm{s}$ the front slows down. In a finite system, the depletion zone moves and extends until it reaches the boundary and the system equilibrates in a state partitioned between a liquid with density $\rho_\mathrm{l}$ and crystal with density $\rho_\mathrm{s}$ with a stationary front between them. For a PFC model the role of the depletion zone in crystal growth is discussed in Ref.~\cite{TGTD2011sm}.

The speed of the hexagons-to-liquid front in the regime $17.0\lesssim \beta\mu\lesssim 19.6$ is determined uniquely (see Appendix). Refs.~\cite{HN00} and \cite{DSSS} predict that this is no longer the case for $\beta \mu \gtrsim 19.6$, but in practice we find that the front has a well-defined speed, possibly as a result of pinning of the stripes-to-liquid front to the stripes behind it, and of the hexagons-to-stripes front to the heterogeneity on either side. Both effects are absent from the amplitude equation formulation employed in Refs.~\cite{HN00} and \cite{DSSS}, that we analyse in the Appendix. Moreover, when the hexagon speed reaches the speed predicted by the marginal stability theory for the stripe state, the two fronts appear to lock and thereafter move together. In the theory based on amplitude equations summarized in the Appendix, the interval of stripes between the two fronts appears to have a unique width, depending on $\beta\mu$, a prediction that is consistent with our DDFT results. We have not observed the ``unlocking'' of the hexagons-to-stripes front from the stripes-to-liquid front noted in Ref.\ \cite{HN00} at yet larger values of $\beta\mu$. Possible reasons for this are discussed in the Appendix.

\begin{figure}
   \centering
   \includegraphics[width=0.9\columnwidth]{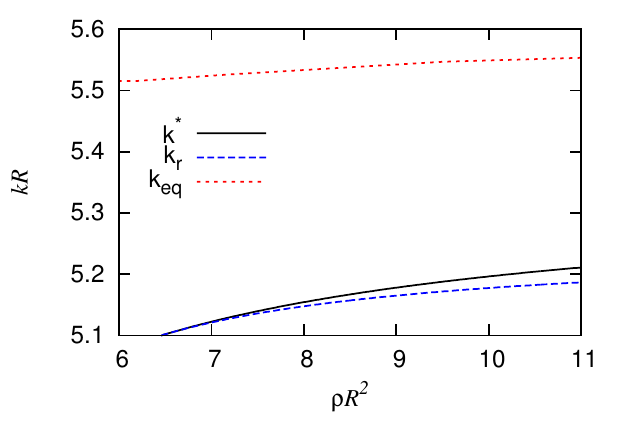} 
   \caption{(Color online) The wave number $k^*$ of the stripe state produced behind the front as a function of density for the GEM-4 fluid with $\beta\epsilon=1$, obtained from Eq.~(\ref{eq:kbehind}) together with the wave numbers $k_{\rm r}$ of the 1D oscillations at the leading edge of the front and $k_{\rm eq}\equiv 2 \pi/\lambda$, where $\lambda$ is the distance between lattice planes in the {\em equilibrium} hexagonal state. This wavelength is very different from the wavelength of the oscillations produced by the advancing front, $2\pi/k^*$.
}
   \label{fig:k_star}
\end{figure}

It is clear, therefore, that the 1D analysis based on Eqs.~(\ref{eq:speed1}) and (\ref{eq:speed2}) allows us to calculate the front speed when the unstable liquid is quenched deeply enough so that fronts propagate via linear processes. In addition to the front speed $c$ this analysis gives $k_{\rm r}$, the wave number of the growing perturbation at the leading edge of the front and $k_{\rm im}$, which defines the spatial decay length of the density oscillations in the forward direction. Within the 1D description the pattern left behind by the front is a large amplitude periodic state with wave number $k^*$, say. When no phase slips take place, this wave number is given by the expression \cite{BBDKL85,ARTK12}
\begin{equation} 
k^*=k_{\rm r}+\frac{1}{c}{\rm Im}[\omega(k)].
\label{eq:kbehind}
\end{equation} 
The wave number $k^*$ differs in general from $k_{\rm r}$. Moreover, as demonstrated in Ref.\ \cite{ARTK12} and confirmed in Fig.\ \ref{fig:k_star} for a GEM-4 crystal with temperature $\beta \epsilon=1$, the wavelength $2\pi/k^*$ of the density modulation that is created by the passage of the front can be very different from the scale $2\pi/k_{\rm eq}$ of the minimum free energy crystal structure which corresponds here to hexagonal coordination. The propagation of the solidification front therefore produces a frustrated structure that leads to the formation of defects and disorder in the crystal. Thus, we identify two sources of frustration: the wave number mismatch and the competition between the stripe state deposited by the advancing front and its subsequent transformation into a 2D hexagonal structure with a different equilibrium wavelength. Both effects generate disorder behind the advancing front and significant rearrangements in the structure of the modulation pattern occur as the system attempts to lower its free energy via a succession of local changes in the wavelength of the density modulation \cite{ARTK12}.

\begin{figure}
   \centering
   \includegraphics[width=0.9\columnwidth]{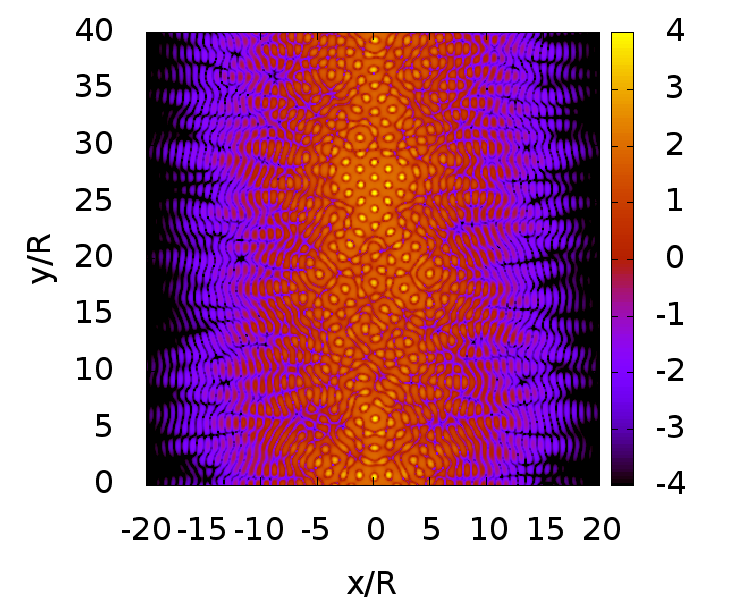}
  \includegraphics[width=0.9\columnwidth]{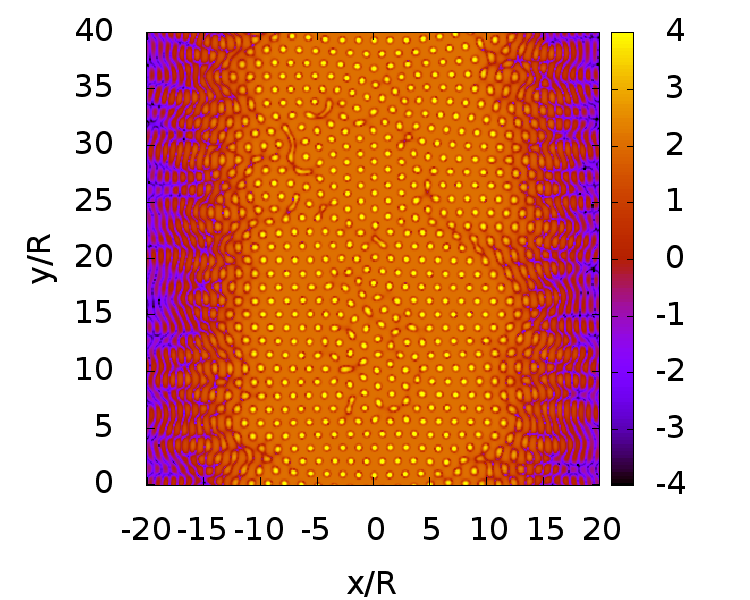}
\caption{(Color online) Density profiles obtained from DDFT for an unstable GEM-4 fluid with bulk density $\rho_0R^2=8$. To facilitate clear portrayal of the front structure we plot the quantity $\ln(R^2|\rho(\rr)-\rho_0|)$. Solidification is initiated along the vertical line $x=0$ at time $t^*=0$. This produces two solidification fronts, one moving to the left, the other to the right, moving away from the line $x=0$. The upper profile is for the time $t^*=1$ and the lower for $t^*=1.4$. We see significant disorder as the front creates density modulations that are not commensurate with the equilibrium crystal structure.}
   \label{fig:front_profiles}
\end{figure}

\begin{figure}

\includegraphics[width=0.9\columnwidth]{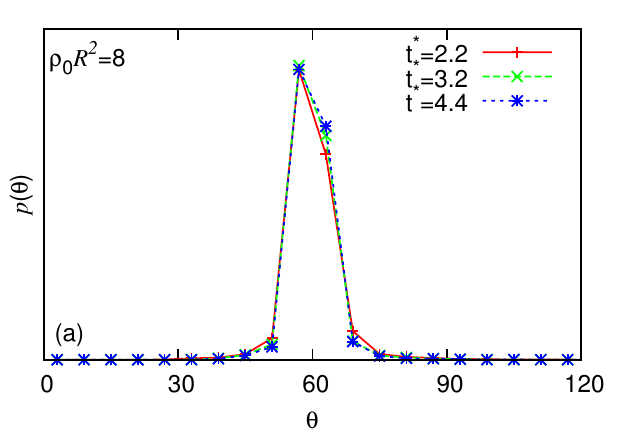} 
 
\includegraphics[width=0.99\columnwidth]{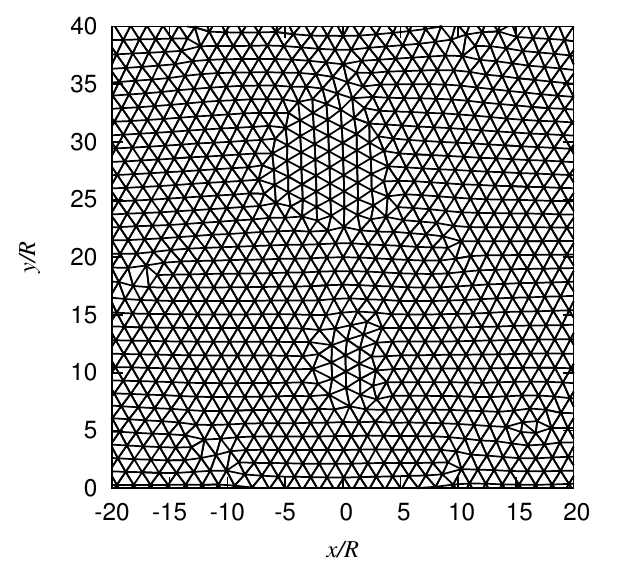}

\includegraphics[width=0.99\columnwidth]{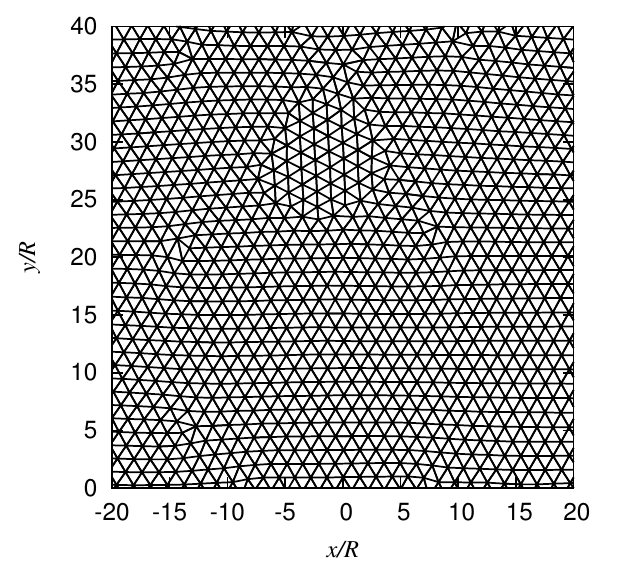}
    
\caption{(Color online) Top panel: the angle distribution $p(\theta)$ at times $t^*=2.2$, 3.2 and 4.4 after the initiation of a solidification front for a GEM-4 fluid with bulk density $\rho_0R^2=8$ (cf.\ Fig.\ \ref{fig:front_profiles}) computed from the triangles of a Delauney triangulation on the density peaks of the profile from DDFT (middle panel: $t^*=2.2$, bottom panel: $t^*=4.4$).}
   \label{fig:Delauney}
\end{figure}

This ageing process can be rather slow \cite{ARTK12}. We illustrate its properties in Figs.~\ref{fig:front_profiles} and \ref{fig:Delauney}. Figure \ref{fig:front_profiles} displays the density profile in a part of the domain as computed from DDFT, and confirms the presence of substantial disorder in the crystalline structure close behind the advancing solidification front. There are actually two fronts in the profiles displayed in Fig.\ \ref{fig:front_profiles}, moving to the left and to the right away from the vertical line $x=0$, where the fronts are initiated at time $t=0$. Although there is substantial disorder close behind the front, further back the crystal has had time to rearrange itself into its equilibrium structure, thereby reducing the free energy. Overall, the process is similar to that observed in the PFC model \cite{ARTK12}. We quantify the rearrangement process using Delauney triangulation \cite{delauney}, as shown in Fig.~\ref{fig:Delauney}. Figure \ref{fig:Delauney}(a) displays the bond angle distribution $p(\theta)$ obtained from Delauney triangulation on the peaks of the density profile at various times after the solidification front was initiated. The distribution $p(\theta)$ has a single peak centered near $60^\circ$, which is not surprising since the triangulation on a hexagonal crystal structure yields equilateral triangles. The initial structure has a significant number of (penta-hepta) defects. Over time, the number of these defects gradually decreases, as shown by the fact that the width of the peak in $p(\theta)$ decreases over time, but the defects never completely disappear. These results show that the one-component GEM-4 system is able to rearrange itself after solidification to form a reasonably well-ordered polycrystalline structure, albeit with defects, but with the equilibrium scale $2\pi/k_{\rm eq}$ present throughout the domain.

\section{Results for a binary system}
\label{sec:bin_system}

\begin{figure}
   \includegraphics[width=1.\columnwidth]{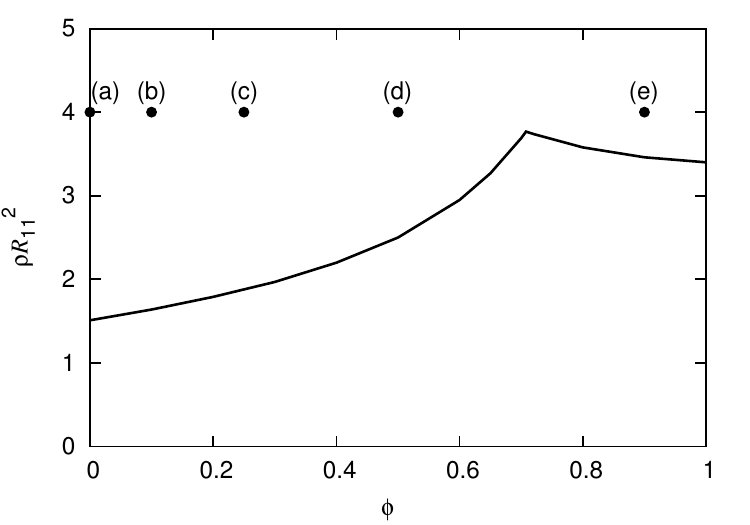} 
   \caption{The linear stability limit for a binary mixture of GEM-8 particles with $\beta \epsilon=1$ and $R_{22}/R_{11}=1.5$ and $R_{12}/R_{11}=1$, plotted in the total density $\rho\equiv\rho_1+\rho_2$ vs concentration $\phi\equiv\rho_1/\rho$ plane. The black circles denote the state points corresponding to the density profiles displayed in Fig.~\ref{fig:binary_mixture_profiles}.}
   \label{fig:lin_stab_line}
\end{figure}

\begin{figure*}
\includegraphics[width=0.65\columnwidth]{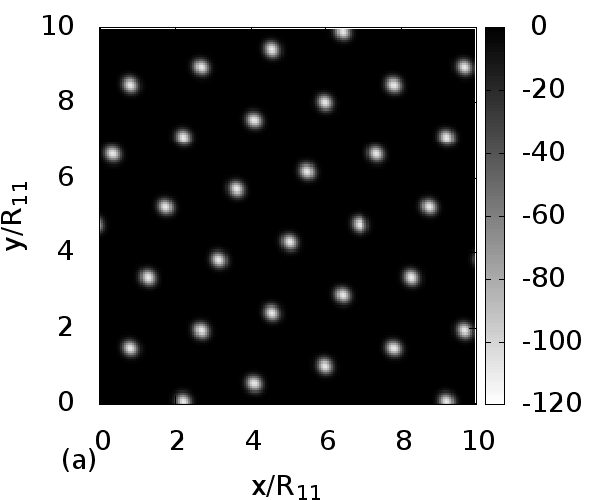}
\includegraphics[width=0.65\columnwidth]{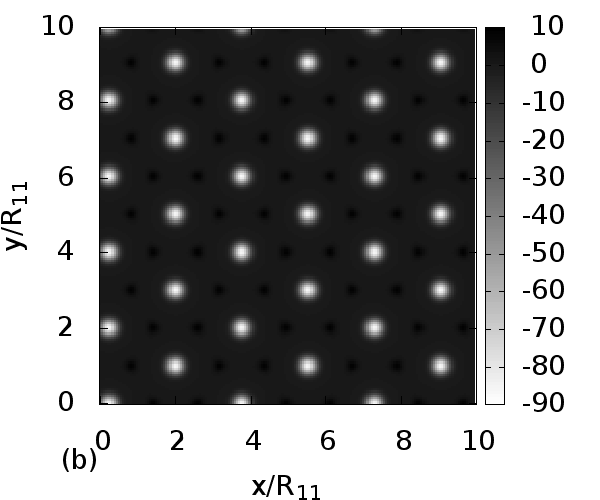}
\includegraphics[width=0.65\columnwidth]{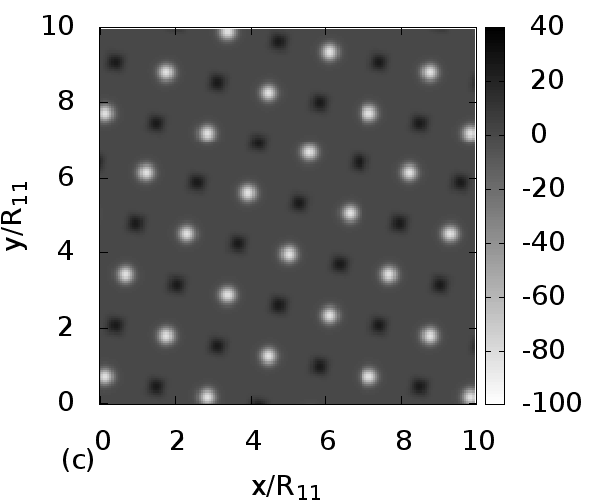}

\includegraphics[width=0.65\columnwidth]{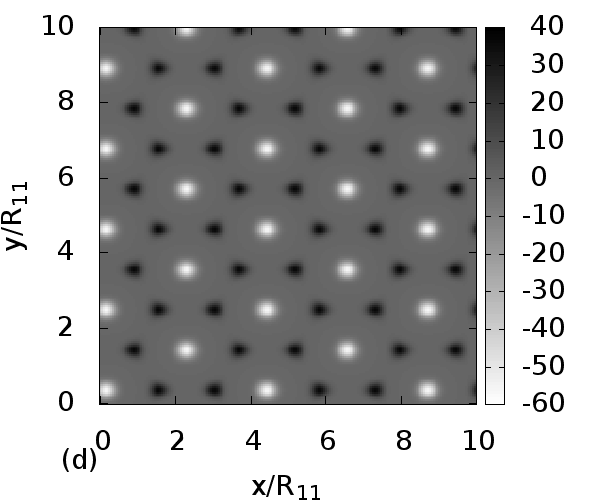}
\includegraphics[width=0.65\columnwidth]{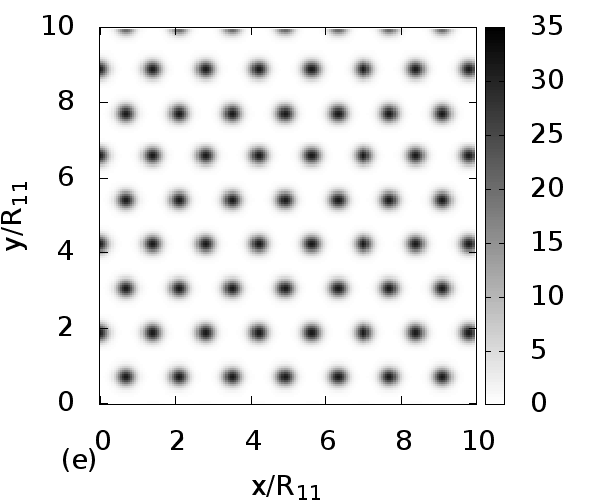}

\caption{Equilibrium crystal structures for a GEM-8 binary mixture with $\beta \epsilon_{ij}=\beta \epsilon=1$ for all $i,j=1,2$, $R_{22}/R_{11}=1.5$, $R_{12}/R_{11}=1$, with total average density $\rho R_{11}^2=4$ and concentrations (a) to (e) $\phi=0$, 0.1, 0.25, 0.5 and 0.9. The structures are shown in terms of the quantity $[\rho_1(\rr)-\rho_2(\rr)]R_{11}^2$, with regions where $\rho_1(\rr)>\rho_2(\rr)$ colored black. All profiles correspond to local minima of the free energy, but we have not checked whether they correspond to global minima at the given state points. We observe a binary square lattice structure in (c), a binary hexagonal lattice structure in (b) and (d), and a simple hexagonal lattice in (a) and (e), where the minority species particles occupy the same lattice sites as the majority species particles, in contrast to the lattice structures in (b)--(d). The density profiles of species 1 or 2 in case (e) are very similar to the profile shown, the only difference being the height of the density peaks.}
   \label{fig:binary_mixture_profiles}
\end{figure*}

\begin{figure}
\includegraphics[width=0.9\columnwidth]{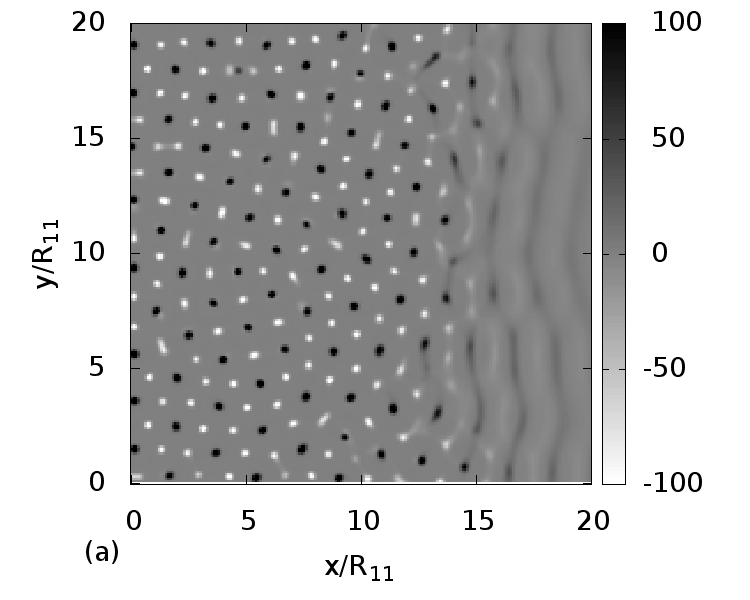}
\includegraphics[width=0.9\columnwidth]{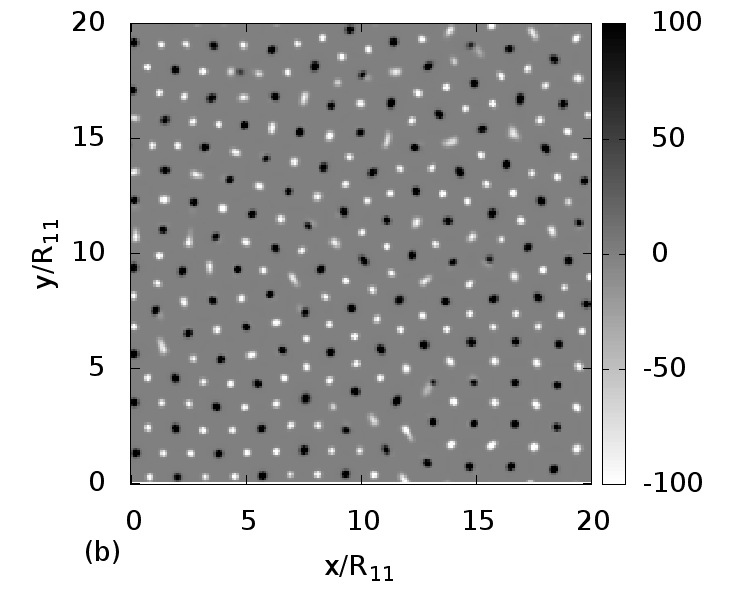}
\caption{Snapshots of a solidification front in a GEM-8 mixture with $\beta \epsilon_{ij}=\beta \epsilon=1$ for all $i,j=1,2$, $R_{22}/R_{11}=1.5$ and $R_{12}/R_{11}=1$, advancing from left to right into an unstable fluid with $\rho R_{11}^2=8$ and $\phi=0.5$, in terms of the quantity $[\rho_1(\rr)-\rho_2(\rr)]R_{11}^2$. Density peaks of species 1 are colored black while the peaks of species 2 are white. The front was initiated at time $t=0$ along the line $x=0$. The top profile corresponds to time $t^*=0.6$ while the lower profile corresponds to $t^*=3$.}
   \label{fig:front_bin}
\end{figure}

\begin{figure*}
   \centering
   \includegraphics[width=0.66\columnwidth]{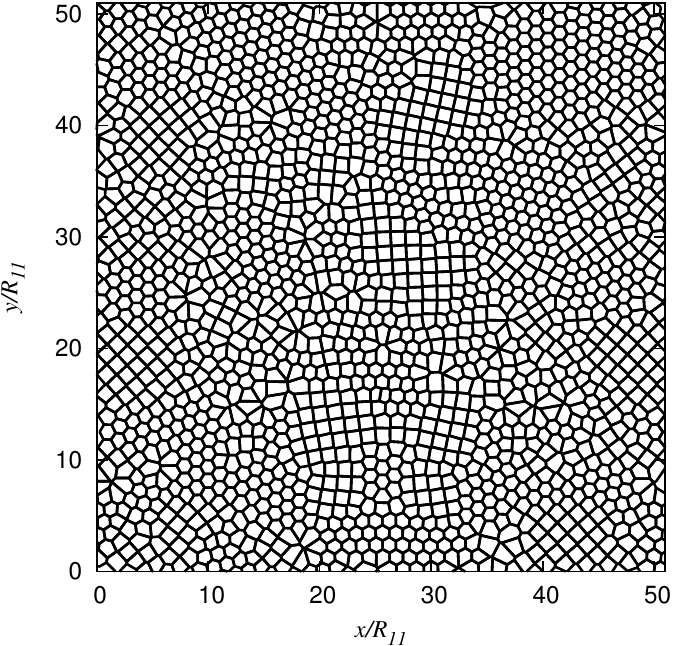}
   \includegraphics[width=0.66\columnwidth]{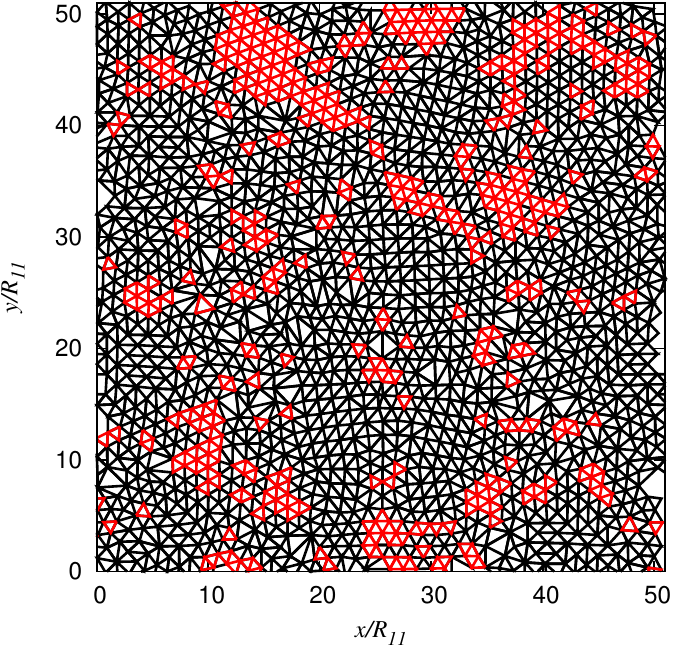}
   \includegraphics[width=0.66\columnwidth]{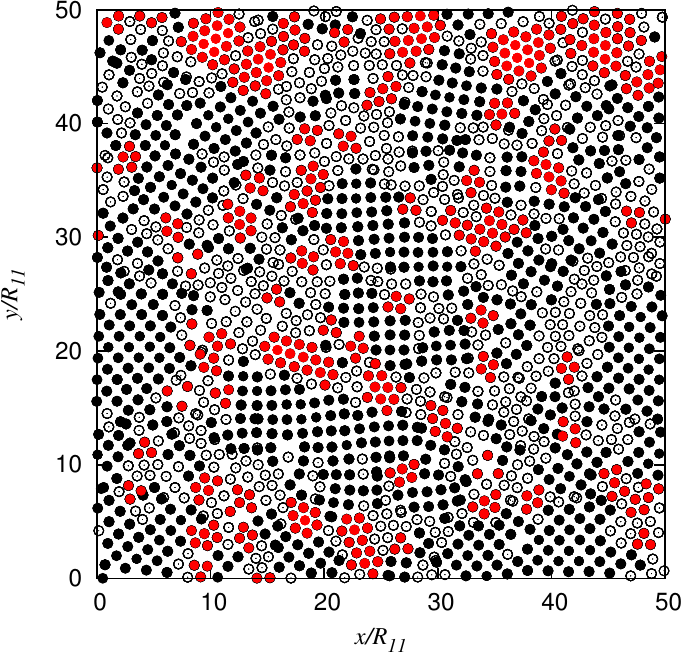}
   
   \includegraphics[width=0.66\columnwidth]{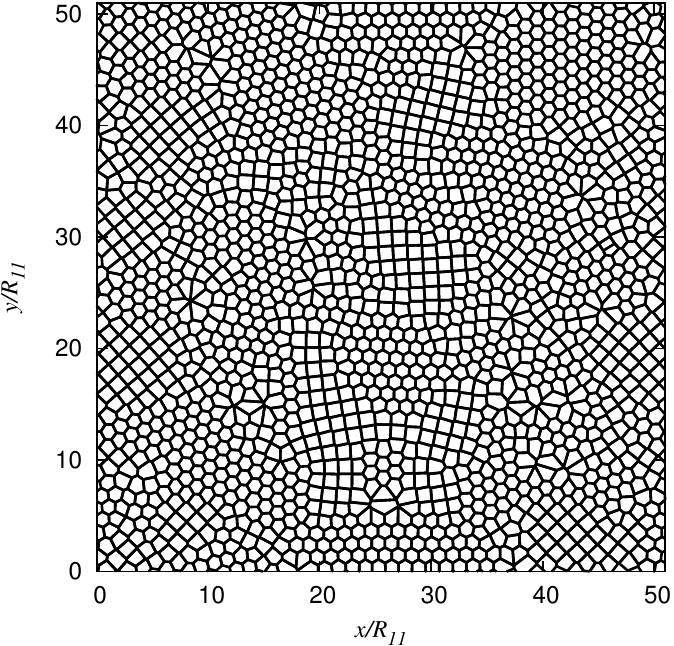} 
   \includegraphics[width=0.66\columnwidth]{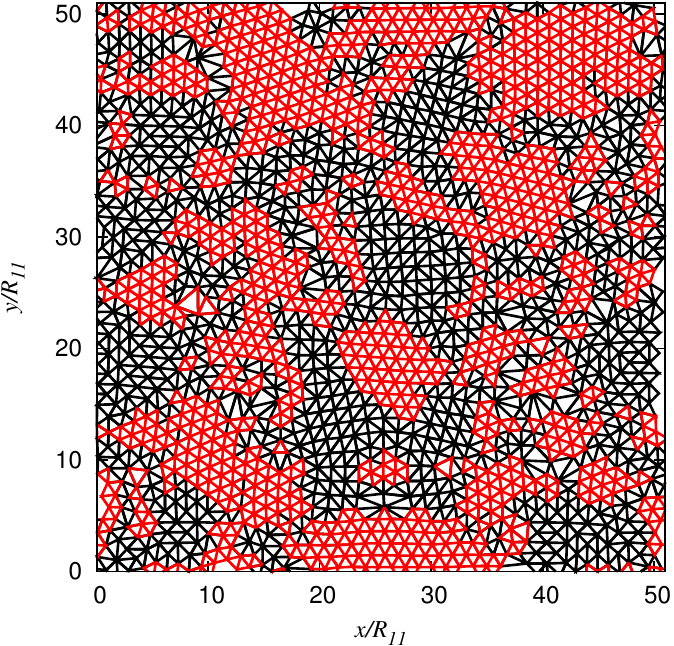} 
   \includegraphics[width=0.66\columnwidth]{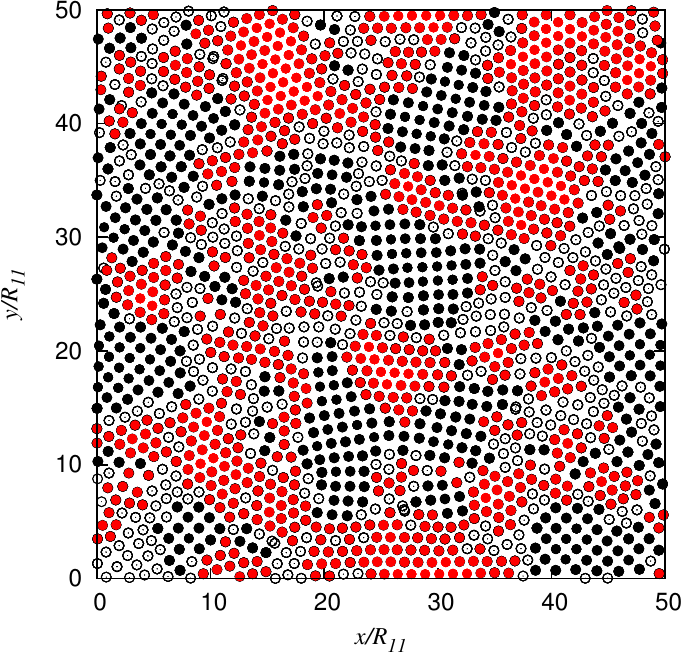}
   
\caption{(Color online) Analysis of the density peaks in the density profile in a GEM-8 mixture with $\beta \epsilon_{ij}=\beta \epsilon=1$ for all $i,j=1,2$, $R_{22}/R_{11}=1.5$ and $R_{12}/R_{11}=1$ and average total density $\rho R_{11}^2=8$ and concentration $\phi=0.5$, formed by a solidification front initiated along the line $x=25$ at time $t=0$. The diagrams along the top row correspond to time $t^*=2$, shortly after the solidification front has exited the domain and before the structure has had time to relax, while the diagrams along the bottom row correspond to time $t^*=400$, when the profiles no longer change in time -- the system has reached a minimum of the free energy. Left: Voronoi diagrams -- the construction reveals the disorder created by the front. The hexagons and squares correspond to the two competing crystal structures. Middle: Delauney triangulation -- domains of the hexagonal phase (equilateral triangles) are highlighted in red, while the remainder, including the right-angled triangles of the square phase, are shown in black. Right: the density maxima are color-coded according to the triangle type they belong to: right-angled triangles are black, equilateral are grey (red online) and scalene are open circles. Comparing the upper to the lower diagrams, we see that over time there is an increase in the size of the domains of the two different crystal structures.}
   \label{fig:Morgan_analysis}
\end{figure*}

\begin{figure}[htbp] 
   \centering
   \includegraphics[width=1.\columnwidth]{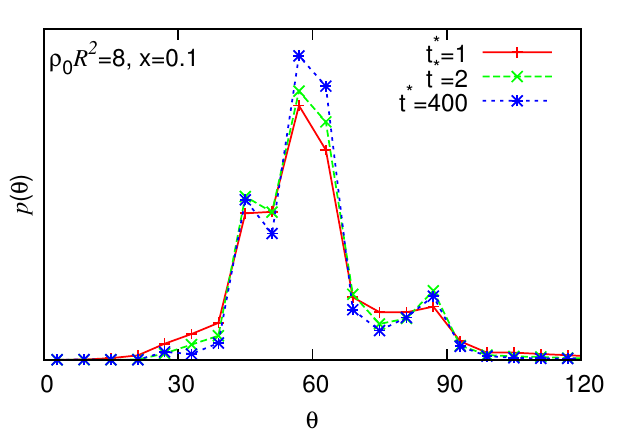}
    \caption{(Color online) Time evolution of the bond angle distribution function from Delauney triangulation, corresponding to the results in Fig.~\ref{fig:Morgan_analysis}.}
   \label{fig:bond_angle_Morgan_analysis}
\end{figure}

Our results from the previous section and also those in \cite{ARTK12} indicate that solidification fronts for systems that have been deeply quenched in general do not produce density modulations with the wavelength of the equilibrium crystal structure. In the quenched one-component fluid discussed in the previous section, the system is subsequently able to rearrange to form the crystal, with only a few defects remaining. However, this begs the interesting question whether in some systems the density peaks are not able to rearrange so that the disorder generated by the solidification front remains. What is well known from the glass transition literature is that quenched binary mixtures are far more likely than one-component systems to form a glass instead of an ordered crystal -- see, for example, Ref.\ \cite{BHHP87}. In order to pursue this idea, we have performed similar computations for a binary mixture of GEM-8 particles with $\beta \epsilon_{ij}=\beta \epsilon=1$ for all $i,j=1,2$ and $R_{22}/R_{11}=1.5$ and $R_{12}/R_{11}=1$. In Fig.\ \ref{fig:lin_stab_line} we display the linear instability threshold for different values of the concentration $\phi\equiv\rho_1/\rho$, where $\rho\equiv\rho_1+\rho_2$ is the total density and $\rho_1$, $\rho_2$ are the densities of the two components of the mixture. For state points above the linear instability threshold line in Fig.\ \ref{fig:lin_stab_line} the uniform fluid is unstable and the system freezes to form a periodic solid. This line is obtained by tracing the locus defined by $D(k_c)=0$, where $D(k)$ is given by Eq.\ \eqref{eq:lin_instab_crit} and $k_c\neq0$ is the wave number at the minimum of $D(k)$ (i.e.\ $\frac{d}{dk}D(k=k_c)=0$). The cusp in the linear instability threshold in Fig.~\ref{fig:lin_stab_line} is a consequence of a crossover from linear instability at one length scale to linear instability at a different lengthscale. At the cusp point, which is at $\rho R_{11}^2=3.77$ and $\phi=0.708$, the system is marginally unstable at two length scales \cite{ARK13}.

This binary mixture exhibits at least four different crystalline phases; examples of these are displayed in Fig.\ \ref{fig:binary_mixture_profiles}. Owing to the fact that the number of potential crystal structures for binary systems of soft-core particles is rather large, we have not attempted to calculate the full phase diagram for this system or the location of the phase transitions between the different structures observed. For clarity the figure shows the quantity $[\rho_1(\rr)-\rho_2(\rr)]R_{11}^2$ with regions where the density of species 1 is higher than that of species 2 indicated in black. For large values of the concentration $\phi$, the system forms a simple hexagonal crystal that is essentially the same as that formed by the pure species 1 system. The minority species 2 particles simply join in low concentration the density peaks formed by the majority species 1 particles -- see Fig.\ \ref{fig:binary_mixture_profiles}(e). Similarly, for very low concentrations $\phi$, the system forms a simple hexagonal crystal, essentially that formed by the pure species 2 system -- see Fig.\ \ref{fig:binary_mixture_profiles}(a). However, for intermediate densities the system forms a binary hexagonal crystal structure, where the two different particle species sit on different lattice sites. Examples of this crystal structure are displayed in Figs.\ \ref{fig:binary_mixture_profiles}(b) and (d). We also observe a square crystal structure -- see Fig.\ \ref{fig:binary_mixture_profiles}(c) -- in which the two different species also reside on different lattice sites.

When the system contains roughly the same number of each species of particles, i.e.\ $\phi\approx0.5$, we find that either the square or the binary hexagonal crystal structures can be formed, depending on initial conditions, indicating that there is close competition between these two different crystal structures. This can also be seen in Fig.\ \ref{fig:front_bin}, where we display profiles calculated from DDFT after the uniform fluid is quenched to this state point and a solidification front is initiated along the line $x=0$. These profiles reveal that the front generates regions of both square and hexagonal crystalline structures. Furthermore, the system is highly disordered, as one might expect based on the demonstration in Sec.\ \ref{sec:1comp_fronts} that the density modulations created behind a solidification front in a deeply quenched system do not have the same wavelength as the equilibrium crystal. Thus, significant rearrangements are needed to get to the equilibrium structure. In the present case, there are two competing structures (squares and hexagons) and the resulting profile contains a mixture of the two. However, because the system is a binary mixture, it is unable to rearrange over time and so significant disorder remains indefinitely. In Figs.\ \ref{fig:Morgan_analysis} and \ref{fig:bond_angle_Morgan_analysis} we display a more detailed analysis of the structure created by the solidification front, and how this structure evolves over time. This analysis is based on performing a Delauney triangulation on the structures that are formed and determining its dual, the Voronoi diagram \cite{delauney}. To do this we first calculate the locations of all the peaks in the total density profile $\rho(\rr)\equiv\rho_1(\rr)+\rho_2(\rr)$. We include all maxima where the density at the maximum point is $>50R_{11}^{-2}$, and construct the Delauney triangulation and the Voronoi diagram on this set of points. The Voronoi diagrams are displayed on the left in Fig.~\ref{fig:Morgan_analysis} while the center panels display the Delauney triangulation. The upper diagrams correspond to a short time $t^*=2$ after the front was initiated along a line down the centre of the system while the lower profiles correspond to a much later time, $t^*=400$, which is roughly when the structure ceases to evolve in time. In the Voronoi diagram we observe regions of both squares and hexagons and in between these different regions we see various different polyhedra corresponding to the defects along the (grain) boundaries between the regions of different crystal structure and/or orientation. These different crystal regions can also be observed in the Delauney triangulation as regions made up of equilateral triangles (coloured red online), corresponding to the hexagonal structure, and regions of right-angled triangles, corresponding to the square crystal structure. The boundaries between these regions contain scalene triangles. In the right hand panels in Fig.\ \ref{fig:Morgan_analysis} we display the density maxima in $\rho(\rr)$. These are color-coded according to the nature of the local crystal structure around that point. The square crystal regions are displayed as black circles, the hexagonal regions as grey circles (red online) and the density peaks with neither square nor hexagonal local coordination are plotted as open circles. The criteria for deciding to which subset a given density peak belongs is based on the Delauney triangulation: any given triangle with corner angles $\theta_1$, $\theta_2$ and $\theta_3$ is defined as equilateral if $|\theta_i-\theta_j|<5^{\circ}$ for all pairs $i,j=1,2,3$. The vertices of these triangles are colored black. Similarly, triangles are defined as right-angled if for the largest angle $\theta_1$ we have $|\theta_1-90^\circ |<5^{\circ}$ AND for the other two angles $|\theta_2-\theta_3|<5^{\circ}$. The vertices of these triangles are colored grey (red online). The remaining vertices which fall into neither of these categories are displayed as open circles. We see that there are roughly equal-sized regions of both square and hexagonal ordering. The typical size of these different regions increases with the elapsed time after the solidification front has passed through the system. Likewise, the number of maxima that do not belong to either crystal structure (open circles) decreases with elapsed time, as the system seeks to minimize its free energy.

In Fig.\ \ref{fig:bond_angle_Morgan_analysis} we plot the distribution function $p(\theta)$ for the different bond angles obtained from Delauney triangulation, for three different times after the initiation of the solidification front. It has three maxima: one near 45$^\circ$, another at 60$^\circ$ and the other near 90$^\circ$. The peak at 60$^\circ$ is the contribution from the regions of hexagonal ordering (equilateral triangles) and the two peaks at 45$^\circ$ and 90$^\circ$ come from the regions of square ordering (right-angled triangles in the Delauney triangulation). The peak at 45$^\circ$ is, of course, twice as high as the peak at 90$^\circ$. We also observe that the peaks are much broader at short times, $t^*=1$, $2$, after the solidification front was initiated, than in the final structure from time $t^*=400$. These results provide an indication of the degree of disorder and number of defects in the system; the fact that the peaks become sharper over time is a consequence of the fact that the amount of disorder in the system decreases over time. Nonetheless, the peaks in $p(\theta)$ are still rather broad in the final state, indicating that significant strain and disorder remain in the structure.

\section{Concluding remarks}
\label{sec:conc}

In this paper we have seen that a deep quench generates a solidification front whose speed is correctly predicted from the dispersion relation using the marginal stability Ansatz. The front leaves behind a nonequilibrium crystalline state with many defects and a characteristic scale that differs substantially from the wavelength of the crystal in thermodynamic equilibrium. Subsequent aging generates domains with different orientations but in one-component systems the number of defects continues to decrease over time. In two-component systems different crystalline phases may compete, providing an additional source of disorder in the system, and the minority species may block rearrangement of the particles, thereby freezing the disorder in place, and leaving an amorphous solid with glass-like structure.

When the quench is shallow, the speed of the solidification front is slow and the amount of disorder generated by its passage is reduced. However, in this regime the front speed in a 2D system is no longer correctly predicted by the 1D marginal stability condition because the front becomes a pushed front, i.e., its speed is determined by nonlinear processes. As a result the speed becomes an eigenvalue of a nonlinear eigenvalue problem as summarized in the Appendix. The solution of this problem reproduces the qualitative features of Fig.~\ref{fig:speed} computed from numerical simulations of the DDFT for a one-component GEM-4 system (see Appendix), thereby providing support for this interpretation of Fig.~\ref{fig:speed}.

In particular, in the region of the phase diagram where the liquid is linearly stable and solidification fronts propagate via nonlinear processes, solidification must be nucleated -- a process that requires the system to surmount a free energy barrier. Once initiated, the resulting solidification front generates disorder in the system by the processes discussed above. However, in addition to these the nucleation process itself may play an important role as discussed in Refs.~\cite{TPTTG11, BaKl13, TXX14, GrTo14}. These studies show that the critical nucleus is likewise a structure that may be incommensurate with the equilibrium crystal structure so that the nucleation process itself can generate disorder in the system. This is especially so as one approaches the linear stability threshold, where the critical nucleus is predicted to have an `onion'--like structure \cite{BaKl13}. The second shell of the `onion' is incompatible with the equilibrium crystal structure, potentially leading to the growth of an amorphous phase, a suggestion supported by recent experimental results \cite{TXX14, GrTo14}. While one-component systems may subsequently be able to rearrange to form a well-ordered crystal, binary systems appear unable to escape the resulting disordered structure.

In the present work, we have studied solidification using DDFT with solidification initiated along a straight line (cf.\ Figs.~\ref{fig:front_profile} and \ref{fig:front_profiles}). The resulting fronts are straight, enabling us to study the front speed and wavenumber selection. For example, the fronts in the linearly unstable liquid in Fig.\ \ref{fig:front_profiles} are initiated by adding a small zero-mean random perturbation along the line $x=0$ to the initially uniform density profile. In reality, however, solidification fronts are initiated throughout the system at random locations, determined by the fluctuations in the system. This is equivalent to initiating fronts simultaneously at many points in the system. These fronts then propagate through the system, colliding and interacting, leading to the formation of the solid phase. To model this process, we add a small zero-mean random perturbation to the initial density profile at all points in the system. The final $t\to\infty$ density profiles produced in this way (not displayed) are very similar to those produced by initiating the solidification front along a single line. If instead of DDFT we employed kinetic Monte Carlo, or Brownian dynamics or even molecular dynamics computer simulations to study solidifation in systems of particles interacting via the potentials in Eq.\ \eqref{eq:pair_pot}, we would first equilibrate the system in the liquid phase at a higher temperature and then quench to a temperature where a solid forms. The dynamics following such a quench is very similar to that predicted by DDFT from an initial density profile with random noise at all points in the system, as is the case for the related soft-core fluid model discussed in Ref.\ \cite{ARK13}. We are thus confident that DDFT gives an excellent description of the system.

We mention, finally, that the behavior of the 2D PFC model studied in \cite{ARTK12} is qualitatively different from the 2D DDFT model studied here. In the PFC model there is a temperature-like parameter $r<0$, such that $(r+1)/2$ is the coefficient of the $\phi^2$ term in the PFC free energy. For the larger values of $|r|$ considered in \cite{ARTK12}, the linear instability threshold lies within the thermodynamic coexistence region between the liquid phase and the hexagonal crystalline phase. Thus, for these values of $|r|$, the hexagonal phase advances into the liquid at a well-defined speed determined by a linear mechanism as described by the marginal stability analysis. This is in contrast to the present DDFT model where the linear instability boundary lies outside the coexistence region (Fig.~\ref{fig:phase_diag}) and fronts between the hexagonal and liquid phases can propagate with speed determined either by a linear or a nonlinear mechanism, depending on parameters. However, for smaller values of $|r|$ the linear instability line in the PFC model does lie outside the coexistence region \cite{WK07, TP14} and in this case the behavior of the PFC system should be similar to that observed in the present study.

{\bf Acknowledgements}: The work of EK supported in part by the National Science Foundation under grant No. DMS-1211953 and a Chaire d'Excellence Pierre de Fermat of the R\'egion Midi-Pyr\'en\'ees, France. AJA and UT thank the Center of Nonlinear Science (CeNoS) of the University of M\"unster for recent support of their collaboration. We thank the anonymous referees for comments that helped shape the discussion in Sec.\ \ref{sec:conc}.

\section{Appendix: 2D front propagation into an unstable state}
\label{Appendix}

Figure \ref{fig:speed}(b) shows the front velocity $c$ as function of the chemical potential $\mu$ as computed from direct numerical simulations of a GEM-4 fluid with temperature $k_BT/\epsilon=1$ and compares the result with the prediction of the marginal stability calculation reported above (red solid line). The latter agrees well with the measured speed for larger values of $\mu$ but there is a substantial disagreement near threshold.

\begin{figure}[t] 
\includegraphics[width=0.9\hsize]{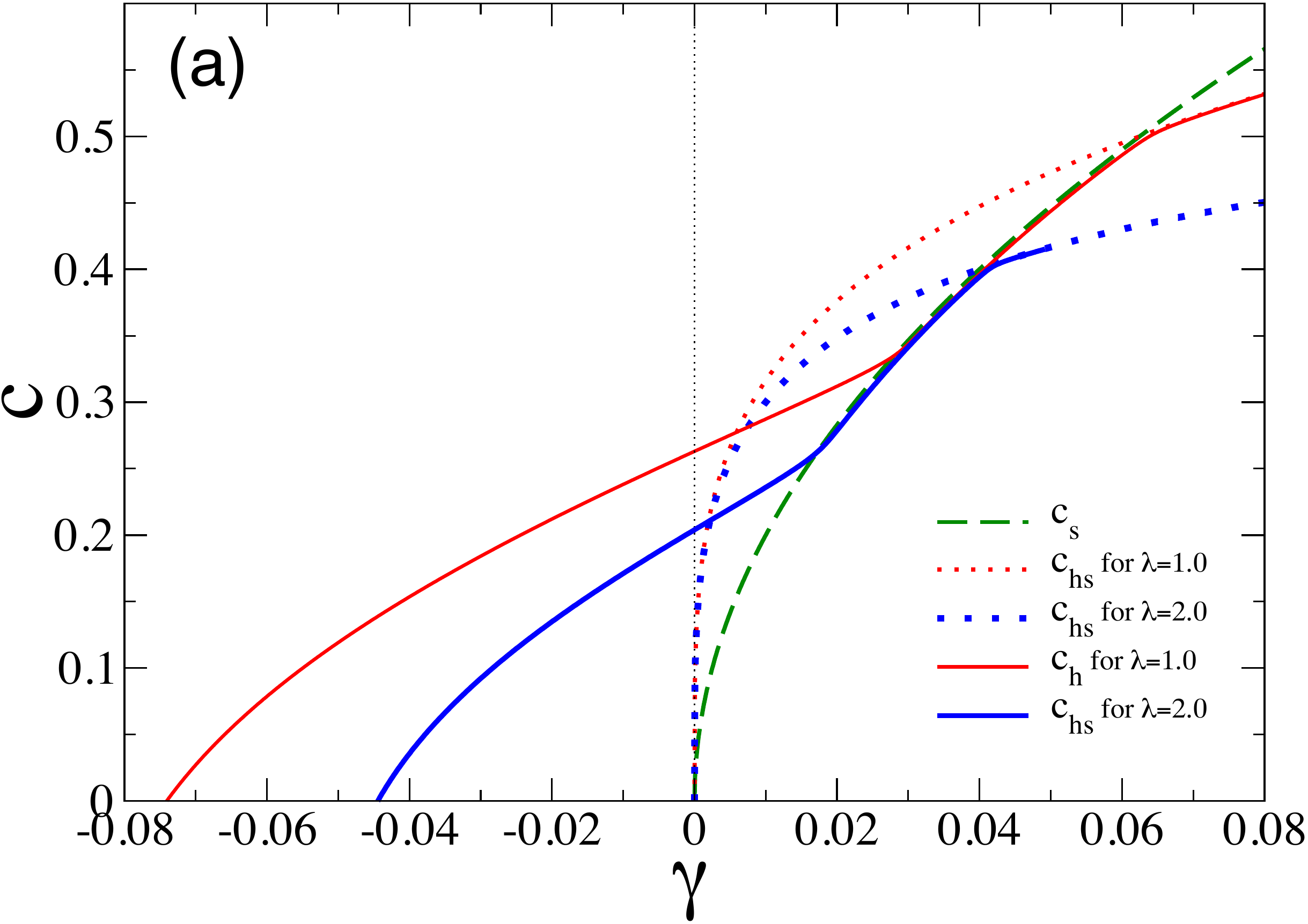}
\includegraphics[width=0.9\hsize]{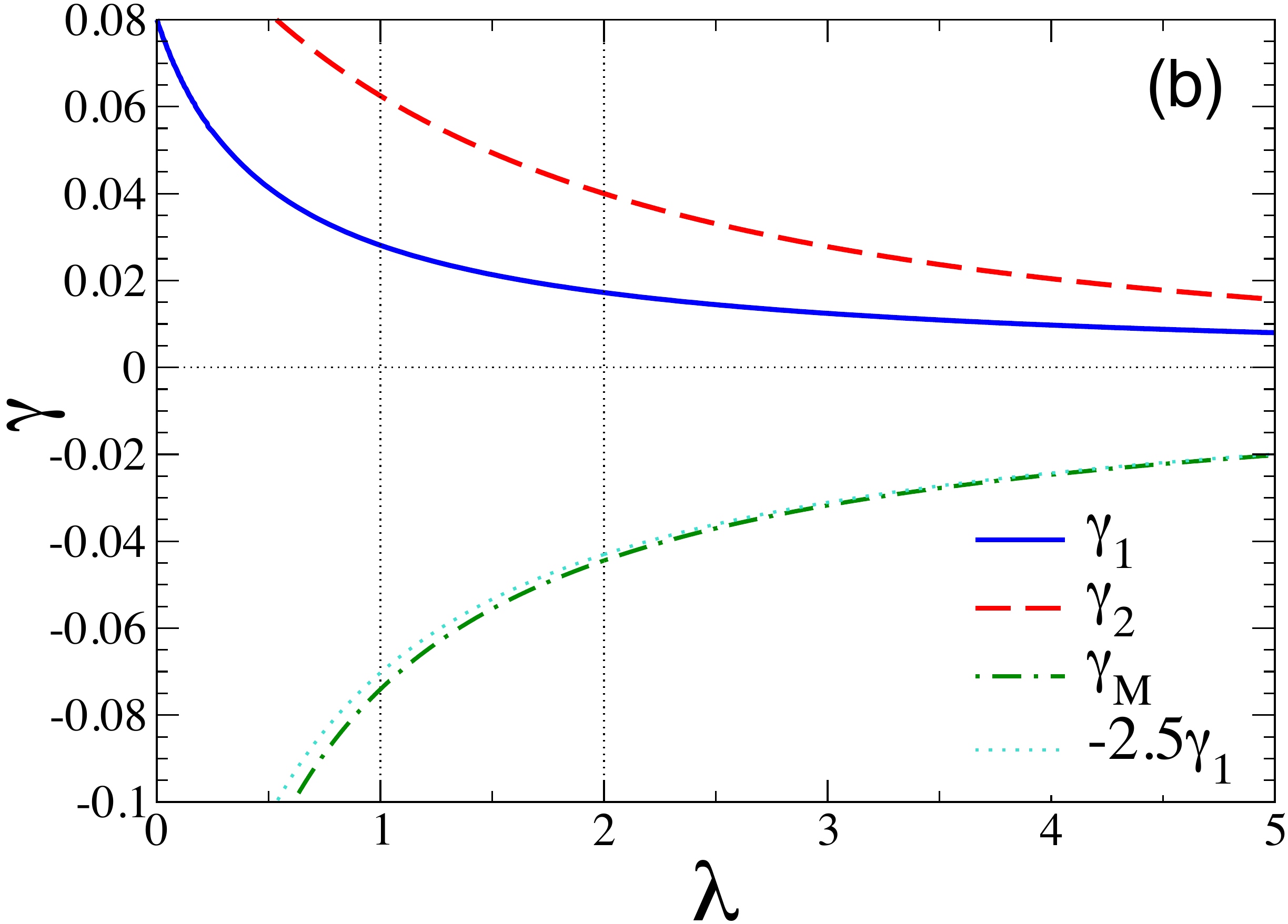}
    \caption{(Color online) (a) The speeds $c_{\rm s}$, $c_{\rm hs}$ and $c_{\rm h}$ defined in the text as a function of $\gamma$ computed from the model system (\ref{A})--(\ref{B}) for $\lambda=1$ and $\lambda=2$. The results for $\lambda=1$ agree with those in Ref.~\cite{HN00}. The full range of values of $\gamma$ is shown including the Maxwell points $\gamma_{\rm M}$, where $c=0$, and the location of the critical values $\gamma_1$ and $\gamma_2$, where $c_{\rm h}=c_{\rm s}$ and $c_{\rm hs}=c_{\rm s}$, respectively. (b) The location of the Maxwell point and the critical values $\gamma_1$ and $\gamma_2$ as a function of the nonlinear coupling coefficient $\lambda$. The dotted line shows $-2.5\gamma_1$ and indicates that, in the range considered, the ratio $\gamma_M/\gamma_1$ is nearly constant.
}
\label{fig:speedHN}
\end{figure}

The reason for this discrepancy was elucidated by Hari and Nepomnyashchy \cite{HN00}, following earlier work by Csah\'ok and Misbah \cite{CM99}. The results of \cite{HN00} were largely confirmed in subsequent work by Doelman et al \cite{DSSS}. The work of Hari and Nepomnyashchy is based on a detailed study of a set of model equations describing the spatial modulation of a pattern of (small amplitude) hexagons:
\begin{eqnarray}\notag
\frac{\partial A_k}{\partial t}=&\gamma A_k+\frac{\partial^2 A_k}{\partial x_k^2}+A^*_{[k-1]}A^*_{[k+1]}\\
&-(|A_k|^2+\lambda|A_{[k-1]}|^2+\lambda|A_{[k+1]}|^2)A_k,
\end{eqnarray}
for $k=0,1,2$, where the $A_k$ are the complex amplitudes of the three wavevectors ${\bf n}_0\equiv (1,0)k_c$, ${\bf n}_1\equiv (-1,\sqrt{3})k_c/2$, ${\bf n}_2\equiv (-1,-\sqrt{3})k_c/2$ \cite{GSK}, and $x_k\equiv{\bf x}\cdot{\bf n}_k$. Here $k_c$ is the critical wave number at onset of the hexagon-forming instability ($\gamma=0$), and $[k\pm1]\equiv(k\pm1) ({\rm mod} 3)$. These equations constitute a gradient flow with free energy
\begin{eqnarray}
{\cal F}\equiv\int_{-\infty}^{\infty}L(x,t)\,dx,
\end{eqnarray}
where
\begin{eqnarray}
L=\sum_{k=0}^2\frac{1}{2}|\frac{\partial A_k}{\partial x_k}|^2-V
\end{eqnarray}
and
\begin{eqnarray}\notag
V\equiv&&\sum_{k=0}^2\biggl(\frac{1}{2}\gamma|A_k|^2-\frac{1}{4}|A_k|^4\biggr)+A_0^*A_1^*A_2^*\\ \notag
&&-\frac{\lambda}{2}\biggl(|A_0|^2|A_1|^2+|A_1|^2|A_2|^2+|A_2|^2|A_0|^2\biggr).
\end{eqnarray}

We focus on planar fronts perpendicular to ${\bf n}_0\equiv (1,0)k_c$ and thus suppose that the dynamics is independent of the variable $y$ along the front. Symmetry with respect to $y\rightarrow -y$ implies that $A_1=A_2\equiv B$, say. Absorbing the wave number $k_c$ in the variable $x$, and writing $A_0\equiv A$ we obtain the equations
\begin{equation}
\frac{\partial A}{\partial t}=\frac{\partial^2 A}{\partial x^2}+\gamma A+B^2-A^3-2\lambda AB^2 \label{e1}
\end{equation}
\begin{equation}
\frac{\partial B}{\partial t}=\frac{1}{4}\frac{\partial^2 B}{\partial x^2}+\gamma B+AB-(1+\lambda)B^3-\lambda A^2B.\label{e2}
\end{equation}
In writing these equations we have assumed that $A$ and $B$ are real to focus on the behavior of the amplitudes, thereby setting the phase $\Phi\equiv{\rm arg}(A)+2{\rm arg}(B)$ that distinguishes so-called up-hexagons from down-hexagons to zero \cite{GSK}.

These equations have solutions in the form of regular hexagons $(A,B)=(A_{\rm h}^{\pm},A_{\rm h}^{\pm})$, stripes $(A,B)=(A_{\rm s},0)$ and the homogeneous liquid state $(A,B)=(0,0)$, where
\begin{eqnarray}
A_h^{\pm}=\frac{1\pm\sqrt{1+4\gamma(1+2\lambda)}}{2(1+2\lambda)},\qquad A_s=\sqrt{\gamma},
\end{eqnarray}
corresponding to the critical points of the potential $V(A,B)=\frac{1}{2}\gamma(A^2+2B^2)+AB^2-[\frac{1}{4}A^4+\lambda A^2B^2+\frac{1}{2}(1+\lambda)B^4]$. The bifurcation to hexagons at $\gamma=0$ is transcritical and for $\gamma<0$ there are two hexagon branches: an unstable branch of small amplitude hexagons $A_{\rm h}^-$ and a stable branch of large amplitude hexagons $A_{\rm h}^+$. These annihilate at a saddle-node bifurcation at $\gamma=\gamma_{\rm sn}\equiv -\frac{1}{4(1+2\lambda)}$. Note that without loss of generality we have taken $A_{\rm h}^{\pm}$ and $A_{\rm s}$ to be positive since negative values can be compensated for by choosing $\Phi=\pi$, i.e., by an appropriate spatial translation.

The large amplitude hexagons $A_{\rm h}^+$ and the homogeneous state coexist stably in the subcritical regime, $-\frac{1}{4(1+2\lambda)}<\gamma<0$; the liquid state becomes unstable when $\gamma>0$. A front traveling with speed $c$ to the right, connecting $A_{\rm h}^+$ on the left with the liquid state $A=0$ to the right, takes the form
\begin{eqnarray}
A(x,t)={\tilde A}(\xi),\quad B(x,t)={\tilde B}(\xi),\quad \xi\equiv x-ct,
\end{eqnarray}
where
\begin{eqnarray}
\frac{\partial^2{\tilde A}}{\partial \xi^2}+c\frac{\partial{\tilde A}}{\partial \xi}+\gamma {\tilde A}+{\tilde B}^2-{\tilde A}^3-2\lambda {\tilde A}{\tilde B}^2=0,\label{A}\\
\frac{1}{4}\frac{\partial^2 {\tilde B}}{\partial \xi^2}+c\frac{\partial {\tilde B}}{\partial \xi}+\gamma {\tilde B}+{\tilde A}{\tilde B}-(1+\lambda){\tilde B}^3-\lambda {\tilde A}^2{\tilde B}=0 \label{B}
\end{eqnarray}
with the boundary conditions
\begin{eqnarray}\notag
{\tilde A}={\tilde B}=A_h^+ \quad{\rm as}\quad \xi\rightarrow-\infty,\\
{\tilde A}={\tilde B}=0 \quad{\rm as}\quad \xi\rightarrow\infty. 
\end{eqnarray}
The speed $c$ vanishes in the subcritical regime when $\gamma=\gamma_M<0$ defined by the requirement $V(A_{\rm h},A_{\rm h})=V(0,0)=0$ and is positive for $\gamma>\gamma_{\rm M}$ ($V(A_{\rm h},A_{\rm h})<0$) and negative for $\gamma<\gamma_{\rm M}$ ($V(A_{\rm h},A_{\rm h})>0$). An elementary calculation gives $\gamma_{\rm M}=-\frac{2}{9(1+2\lambda)}$; $\gamma_{\rm M}$ thus corresponds to the Maxwell point between the trivial state $(0,0)$ and the hexagonal state $(A_{\rm h}^+,A_{\rm h}^+)$. Note that $\gamma_{\rm M}/\gamma_{\rm sn}=8/9$, independently of the value of $\lambda$. This prediction of the amplitude equations compares well with our numerical results for a GEM-4 mixture for which the chemical potential $\beta\mu_{\rm sn}\approx16.5$ and $\beta\mu_{\rm M}\approx16.8$ while the linear instability threshold corresponds to $\beta\mu_{\rm lin}\approx19.6$. Thus $(\mu_{\rm M}-\mu_{\rm lin})/(\mu_{\rm sn}-\mu_{\rm lin})\approx0.90$, very close to the predicted value $8/9$.

The situation is more complicated in the supercritical regime where $\gamma>0$ because this regime contains supercritical (but unstable!) stripes oriented parallel to the front. As a result one now finds fronts that connect the hexagonal structure to the stripe pattern and the stripe pattern to the liquid state, in addition to the front connecting the hexagonal structure and the (now unstable) liquid state. The marginal stability condition implies that stripes invade the homogeneous state with speed $c_{\rm s}=2\sqrt{\gamma}$, while an analogous calculation shows that the hexagons invade the unstable stripes with speed $c_{\rm hs}=[\sqrt{\gamma}-(\lambda-1)\gamma]^{1/2}$. This speed exceeds $c_{\rm s}$ in the interval $0<\gamma<\gamma_2\equiv(\lambda+3)^{-2}$, i.e., at $\gamma_2$ one has $c_{\rm hs}=c_{\rm s}$. The dependence of the speeds $c_{\rm hs}$ and $c_{\rm s}$ on $\gamma$ is shown in Fig.~\ref{fig:speedHN}(a) for $\lambda=1$ and $\lambda=2$.

It is evident that the speed $c_{\rm s}$ cannot be selected when $\gamma$ is too close to threshold $\gamma=0$ since $c$ must be positive for all $\gamma>\gamma_{\rm M}$. In the spatial dynamics picture of the front one seeks a heteroclinic connection between $(\tilde{A},\tilde{A})=(A_{\rm h},A_{\rm h})$ and $(0,0)$. Near $(0,0)$ we have the asymptotic behavior 
\begin{eqnarray}
\tilde{A}\sim e^{\kappa_A\xi}\qquad \tilde{B}\sim e^{\kappa_B\xi},\qquad {\rm as}\quad \xi\rightarrow\infty,
\end{eqnarray}
where
\begin{eqnarray}
\kappa_A^{\pm}=-\frac{c}{2}\pm\frac{1}{2}\sqrt{c^2-4\gamma},\quad \kappa_B^{\pm}=-2c\pm 2\sqrt{c^2-\gamma}.
\end{eqnarray}
Evidently, for $\gamma<0$ the stable manifold of $(0,0)$ is two-dimensional, and since one expects the heteroclinic to connect to $(0,0)$ along the slow direction one anticipates that the solution will approach $(0,0)$ in the ``$A$" direction, with $\tilde{A}\sim e^{\kappa_A^-\xi}$ as $\xi\rightarrow\infty$. However, as soon as $\gamma>0$ the stable manifold of $(0,0)$ becomes four-dimensional, and the slowest direction is suddenly $\tilde{A}\sim e^{\kappa_B^+\xi}$. Hari and Nepomnyashchy \cite{HN00} solve the problem (\ref{e1})--(\ref{e2}) numerically and find that for $c<2\sqrt{\gamma_1}$ the front speed departs from the prediction $c=c_{\rm s}$ and instead follows a speed $c=c_{\rm h}$ for which the asymptotic behavior of the front continues to be $\tilde{A}\sim e^{\kappa_A^-\xi}$ as $\xi\rightarrow\infty$, thereby providing a smooth connection to the speed computed for $\gamma<0$. We refer to the value of $\gamma$ at which $c_{\rm h}=c_{\rm s}$ as $\gamma=\gamma_{1}$.

Hari and Nepomnyashchy \cite{HN00} also show that in the region $\gamma_{1}<\gamma<\gamma_2$ both the front connecting the hexagonal state to the stripes and the front connecting the stripes to the liquid state travel with the same speed $c_\mathrm{s}$. As a result the width of the stripe region between the hexagons and the liquid state remains constant; in numerical simulations this width was observed to be independent of the initial conditions adopted, despite the nonuniqueness of the overall front solution, and to increase with $\gamma$. Finally, for $\gamma>\gamma_2$ the front speed $c_{\rm s}>c_\mathrm{hs}$ and the front connecting the stripes to the liquid state outruns the hexagons invading the stripes and the width of the stripe interval in front of the hexagons grows without bound. In our models this behavior was not observed.

Figure \ref{fig:speedHN}(a) shows the computed fronts speeds as a function of the bifurcation parameter $\gamma$ for two values of the single nonlinear coupling coefficient $\lambda$ which is unknown for our GEM-4 model. In both cases the results behave qualitatively like those obtained from DDFT of this model system. In particular, we see that the speed $c_{\rm h}$ of the (pushed) hexagons increases monotonically from zero at the Maxwell point $\gamma_{\rm M}<0$ and terminates on the 1D stripe speed $c_{\rm s}$ obtained from the marginal stability at $\gamma=\gamma_1>0$; both $\gamma_{\rm M}$ and $\gamma_1$ decrease in magnitude as $\lambda$ increases (Fig.~\ref{fig:speedHN}(b)) and this is so for the point $\gamma=\gamma_2$ corresponding to the condition $c_{\rm s}=c_{\rm hs}$ as well. We mention that behavior similar to Fig.~\ref{fig:speedHN}(a) occurs even in 1D, provided only that the stripe state bifurcates subcritically before turning around towards larger values of the forcing parameter \cite{CC97}.

However, despite its qualitative success the model system (\ref{A})--(\ref{B}) fails in one key respect: it is not possible to match quantitatively the DDFT results for a shallow quench (Fig.~\ref{fig:speed}(b)) with the predictions of the model (Fig.~\ref{fig:speedHN}(b)). Specifically, the model predicts that $|\gamma_M|/\gamma_1\approx 2.5$ over the entire range of nonlinear coefficients $\lambda$ in Fig.~\ref{fig:speedHN}(b) while Fig.~\ref{fig:speed}(b) indicates that $|\gamma_M|/\gamma_1\approx 1.4$. For smaller $\lambda$ the ratio becomes yet larger. There are several issues that might contribute to this quantitative mismatch. First, the amplitude equations omit the phenomena of locking of the stripes-to-liquid front to the stripes behind the front, and of locking of the hexagons-to-stripes front to the heterogeneity ahead and behind the front. This is a consequence of modeling periodic structures using constant amplitude states, i.e., by spatially homogeneous states, resulting in the absence of the so-called nonadiabatic effects. Second, the amplitude equations are derived for nonconserved systems, while the DDFT system exhibits conserved dynamics. In the latter case we expect the equations for the amplitudes $A$ and $B$ to be coupled to a large scale mode, much as discussed in the work of Refs.~\cite{maco00,TARGK}. These aspects of the problem will be discussed in a future publication.


\end{document}